\documentclass{jfm}

\def\footerflagdefns#1{}%

\usepackage{amsmath}
\usepackage{amsfonts}
\usepackage{amssymb}

\usepackage{graphicx}
\usepackage{nicematrix} 
\usepackage[pdfpagemode={UseOutlines},bookmarks=true,bookmarksopen=true,bookmarksopenlevel=0,bookmarksnumbered=true,hypertexnames=false,colorlinks,linkcolor={blue},citecolor={blue},urlcolor={red},pdfstartview={FitV},unicode,breaklinks=true]{hyperref}
\usepackage{tikz}

\usepackage{nicefrac}

\usepackage[numbers]{natbib}

\usepackage{todonotes}
\usepackage{lscape}

\usepackage{import}

\usepackage{subfig}

\usepackage{booktabs}

\usepackage{rotating}

\usepackage{supertabular}

\usepackage{floatpag}
\floatpagestyle{empty}
\rotfloatpagestyle{empty}

\newcommand{\scal}[2][]{#2_\mathrm{#1}}
\newcommand{\vect}[2][]{\boldsymbol{#2}_\mathrm{#1}}
\newcommand{\tens}[2][]{\boldsymbol{#2}_\mathrm{#1}}

\newcommand{\scali}[2][]{#2_{#1}}
\newcommand{\vecti}[2][]{\boldsymbol{#2}_{#1}}
\newcommand{\tensi}[2][]{\boldsymbol{#2}_{#1}}

\newcommand{\ave}[2][]{\left\langle{#2}\right\rangle_\mathrm{#1}}
\newcommand{\dnum}[2][]{\mbox{#2}_\mathrm{#1}}

\newcommand{\order}[1]{\ensuremath{\mathcal{O}\left(#1\right)}}
\newcommand{\cross}{\times}

\newcommand{\trace}[1]{\ensuremath{\mathrm{tr}\left(#1\right)}}

\newcommand{\ee}[1]{\ensuremath{\times 10^{#1} }}

\newlength{\explainwidth}

\newlength{\orderunderwidth}
\newcommand{\orderunder}[2]{\settowidth{\orderunderwidth}{\order{#2}} \ensuremath{\underbrace{ \Big . #1}_{\parbox{\orderunderwidth}{\centering\order{#2}}}}}
\newcommand{\uf}{\vect[f]{u}}
\newcommand{\uvectrho}{\vecti[\rho]{u}}
\newcommand{\uslip}{\vect[slip]{u}}

\newcommand{\up}{\vect{u}^p}
\newcommand{\upp}{\vect[p]{u}}
\newcommand{\us}{\vect[s]{u}}
\newcommand{\usp}{\vect[s]{u}(\xp)}
\newcommand{\usq}{\vect[s]{u}(\xq)}
\newcommand{\ui}{\vecti[i]{u}}
\newcommand{\um}{\vect[m]{u}}
\newcommand{\uvect}{\vect{u}}

\newcommand{\bp}{\vect{b}^p}
\newcommand{\Up}{\vect{U}^p}
\newcommand{\Omegap}{\vect{\Omega}^p}
\newcommand{\Ep}{\tens{E}^p}
\newcommand{\Kp}{\tens{K}^p}
\newcommand{\Pp}{P^p}
\newcommand{\tfixp}{\vect[fix]{t}^p}
\newcommand{\ttotp}{\vect[tot]{t}^p}
\newcommand{\tintp}{\vect[int]{t}^p}
\newcommand{\rfix}{\vect{z}}
\newcommand{\lambdafixp}{\scal[fix]{\lambda}^p}
\newcommand{\lambdafix}{\scal[fix]{\lambda}}
\newcommand{\lambdaint}{\scal[int]{\lambda}}
\newcommand{\Itens}{\tens{I}}
\newcommand{\ItensD}{\tens[D]{I}}
\newcommand{\gammap}{\tens{\gamma}^p}
\newcommand{\thetap}{\tens{\theta}^p}
\newcommand{\taup}{\tens{\tau}^p}
\newcommand{\phif}{\scal[f]{\phi}}
\newcommand{\phis}{\scal[s]{\phi}}

\newcommand{\phii}{\scali[i]{\phi}}
\newcommand{\rhof}{\scal[f]{\rho}}
\newcommand{\rhos}{\scal[s]{\rho}}
\newcommand{\rhom}{\scal[m]{\rho}}
\newcommand{\rhoi}{\scali[i]{\rho}}
\newcommand{\pf}{\scal[f]{p}}
\newcommand{\muf}{\scal[f]{\mu}}

\newcommand{\n}{\vect{n}} 
\newcommand{\np}{\vect{n}^p} 
\newcommand{\vnabla}[1][]{\boldsymbol{\nabla}_{#1}}
\newcommand{\vnablax}{\vnabla[\x]{}}
\newcommand{\vnablaxp}{\vnabla[\xp]{}}
\newcommand{\Vf}{\scal[f]{V}}
\newcommand{\Vs}{\scal[s]{V}}
\newcommand{\Vp}{V^p}
\newcommand{\x}{\vect{x}}
\newcommand{\y}{\vect{y}}
\newcommand{\xp}{\vect{x}^p}
\newcommand{\omegap}{\vect{\omega}^p}
\newcommand{\omegas}{\vect[s]{\omega}}
\newcommand{\omegaslip}{\vect[slip]{\omega}}
\newcommand{\omegaf}{\vect[f]{\omega}}
\newcommand{\xq}{\vect{x}^q}
\newcommand{\xqp}{\vect{x}^{qp}}
\newcommand{\Deltaxqp}{\Delta\xqp}

\newcommand{\g}{\vect{g}}
\newcommand{\fp}{f^p}
\newcommand{\rhat}{\vect{d}}
\newcommand{\epsf}{\vect[f]{\epsilon}}
\newcommand{\epss}{\vect[s]{\epsilon}}

\newcommand{\intr}{\int_{r>2a}}
\newcommand{\dr}{d\vect{r}}
\newcommand{\intSp}{\int_{S^p}}
\newcommand{\dSn}{d \scali[\n]{S}}
\newcommand{\intSn}{\int_{\scali[\n]{S}}}

\newcommand{\gammadot}{\dot{\gamma}}
\newcommand{\gammahatdot}{\hat{\gammadot}}
\newcommand{\tensgammadot}{\tens{\gammadot}}
\newcommand{\tensgammadots}{\tens[s]{\gammadot}}
\newcommand{\tensgammadotf}{\tens[f]{\gammadot}}
\newcommand{\tensgammadotm}{\tens[m]{\gammadot}}
\newcommand{\tensgammadoti}{\tensi[i]{\gammadot}}

\newcommand{\tensgammadotbin}{\tens[bin]{\gammadot}}

\newcommand{\tensgammahatdots}{\tens[s]{\gammahatdot}}

\newcommand{\gammadots}{\scal[s]{\gammadot}}

\newcommand{\gammadotsphs}{\scal[sph,s]{\gammadot}}

\newcommand{\gammadotsphi}{\scali[\text{sph},i]{\gammadot}}
\newcommand{\gammadotsphf}{\scal[sph,f]{\gammadot}}
\newcommand{\gammadotsph}{\scal[sph]{\gammadot}}

\newcommand{\chat}[1]{\hat{C}_{#1}}
\newcommand{\dhat}[1]{\hat{D}_{#1}}

  \newcommand{\tauints}{\tens[int,s]{\tau}}
  \newcommand{\muints}{\scal[int,s]{\hat{\mu}}}
  
  \newcommand{\musidones}{\scal[sid1,s]{\hat{\mu}}}
  \newcommand{\musidtwos}{\scal[sid2,s]{\hat{\mu}}}
  
  \newcommand{\mubulkones}{\scal[bulk1,s]{\hat{\mu}}}
  \newcommand{\mubulktwos}{\scal[bulk2,s]{\hat{\mu}}}

  \newcommand{\taudilm}{\tens[dil,m]{\tau}}
  \newcommand{\tauintm}{\tens[int,m]{\tau}}
  \newcommand{\muintm}{\scal[int,m]{\hat{\mu}}}
  
  \newcommand{\musidonem}{\scal[sid1,m]{\hat{\mu}}}
  \newcommand{\musidtwom}{\scal[sid2,m]{\hat{\mu}}}
  
  \newcommand{\mubulkonem}{\scal[bulk1,m]{\hat{\mu}}}
  \newcommand{\mubulktwom}{\scal[bulk2,m]{\hat{\mu}}}

  \newcommand{\tauinti}{\tensi[\text{int},i]{\tau}}
  \newcommand{\muinti}{\scali[\text{int},i]{\hat{\mu}}}
  
  \newcommand{\musidonei}{\scali[\text{sid1},i]{\hat{\mu}}}
  \newcommand{\musidtwoi}{\scali[\text{sid2},i]{\hat{\mu}}}
  
  \newcommand{\mubulkonei}{\scali[\text{bulk1},i]{\hat{\mu}}}
  \newcommand{\mubulktwoi}{\scali[\text{bulk2},i]{\hat{\mu}}}

  \newcommand{\rinfty}{{r_\infty}}

\begin{document}

\title{Multifluid tensorial equations for the flow of semi-dilute monodisperse suspensions}

\shortauthor{D.J.E. Harvie}
\author{Dalton J. E. Harvie\aff{1} \corresp{\email{daltonh@unimelb.edu.au}}}
\affiliation{\aff{1}Department of Chemical Engineering, The University of Melbourne, Parkville, Victoria 3010, Australia}

\maketitle

\begin{abstract}
Using the volume averaging technique of \citet{jackson97}, we derive a set of two-fluid equations that describe the dynamics of a mono-disperse non-Brownian colloidal suspension in the semi-dilute regime.  The equations are tensorial and can be applied in arbitrary geometries.  Closure models are developed that represent the stress surrounding each particle as a sum of stresses due to fluid movement through a fixed bed of particles, and those due to interactions between particles.  Emphasising pragmatism, the developed closure models are consistent with current knowledge of particle interactions in these systems but employ parameters that can be tuned to represent the microstructure of specific particle suspensions.  Within the interaction model, a model for the particle distribution around each particle is used that depends on the strain rate field, allowing anisotropy of the microstructure (and hence normal suspension stresses) to develop within the suspension in response to arbitrary strain fields.  Force moments acting on particles during particle interactions are calculating by summing hydrodynamic contributions between particle pairs, but adjusted to recognise that multi-particle interactions can increase the effective stress generated during each interaction.  Finally, an order of magnitude analysis is performed on the derived momentum equations to determine what terms are significant, leaving only terms in the final equations that are required to predict behaviour during laminar flow within the targeted semi-dilute regime.  In a companion paper (referred to as paper II) \citep{noori24a} we chose sets of microstructure parameters that predict experimentally measured bulk suspension behaviour, creating a link between particle properties (i.e. roughness), suspension microstructure and shear-induced particle migration in arbitrary flow fields.
\end{abstract}
\tableofcontents



\begin{center}
\textbf{Nomenclature}
\end{center}

\begin{supertabular}{p{2cm}p{11cm}}
$a$ & radius of particle \\
$\vect{d}$ & unit vector along line of intersecting particles \\
$\scali[i]{f}^*$ & PDF microstructure parameters \\
$g$ & radial weighting function \\
$\vect{g}$ & acceleration due to gravity \\
$l$ & lengthscale (effective radius) of radial weighting function $g$\\
$L$ & lengthscale of macroscopic flow\\
$\lambdafix$ & increase in fixed stresses due to multi-particle effects\\
$\lambdaint$ & increase in interaction stresses due to multi-particle effects\\
$\muf$ & viscosity of fluid phase\\
$\hat{\mu}$ & interaction viscosity coefficients\\
$\nu$ & volume of a single particle\\
$n$ & number density of particles \\
$\np$ & outward facing unit normal from particle $p$\\
$p$ & non-dimensional microstructure PDF\\
$\pf$ & average pressure within fluid phase\\
$\phii$ & average volume fraction of $i$th phase, $i=$ f or s\\
$\rhoi$ & density of $i$th phase, $i=$ f or s\\
$\rhom$ & density of the mixture\\
$S^p$ & surface of particle $p$\\
$\vect{t}$ & surface traction\\
$\ui$ & average velocity of $i$th phase, $i=$ f or s\\
$\um$ & volume averaged velocity (mixture)\\
$\vecti[\rho]{u}$ & mass averaged velocity\\
$\up$ & velocity of particle $p$\\
$\Vf$ & volume of fluid phase\\
$\Vs$ & volume of solid phase\\
$V$ & total volume of both phases\\
$\Vp$ & volume of particle $p$\\
$\xp, \xq,$ etc & centroid of particle $p$, $q$, etc\\
$\xqp$ & mid-point between particles $p$ and $q$\\
$\Deltaxqp$ & displacement of particle $q$ from $p$\\
$\tensgammadot$ & strain rate tensors\\
$\tens{\tau}$ & suspension level viscous stresses\\
& \\
\multicolumn{2}{l}{\emph{Superscripts}} \\
$p,q,$ etc & represents a property of particle $p$, $q$ etc \\
& \\
\multicolumn{2}{l}{\emph{Subscripts}} \\
f & represents a fluid-phase property\\
p & represents a particle-phase property\\
s & represents a solid-phase property\\
m & represents a non-phase specific property (mixture average)\\
\end{supertabular}

\section{Introduction}

Solid-liquid suspension flows occur in wide variety of industrial, biological and environmental applications.  As discussed in two recent and comprehensive reviews\citep{guazzelli18,morris20a}, despite the large amount of work has been done to understand these complex systems, we do not have the tools available to describe the flow of non-dilute suspensions moving through complex geometries.  This hampers our ability to (e.g.) design flowing particulate technology to suit specific applications.  
Hence, the aim of this paper is to use a methodical averaging approach, combined with physics-informed closure models, to derive a frame-invariant tensorial set of constitutive equations that can predict the flow of mono-disperse non-Brownian colloidal suspensions of spheres within an incompressible and Newtonian fluid, above the dilute concentration limit, and through complex geometries.

Previous works that are relevant to this study come from a variety of research domains.  Considering dilute suspensions where particles interact only with the fluid and not each other, various averaging techniques have been used to derive frame-invariant constitutive equation sets.  Using the same volume averaging technique as employed in the present study, \citet{anderson67,jackson97,jackson98} obtained a set of `two-fluid' equations applicable to a dilute suspension of mono-disperse spheres, with particle stresses modelled as those on an isolated sphere moving in a quadratic velocity field.  By definition `multi-fluid' equations consist of mass and momentum conservation equations for each phase present, with `two-fluid' referring to situations where just a single fluid and single solid phase are present.  \citet{zhang97} derived a very similar set of equations using an ensemble averaging technique.  Both equation sets contain a faxen interphase force, in addition to drag, inertial, viscous and buoyancy terms, and predict a Newtonian suspension with a viscosity that varies linearly with solids volume fraction ($\phis$) as predicted by Einstein ($\muf (1 + \nicefrac{5}{2}\phis)$).  These equations are only suitable for predicting suspension behaviour when the solid volume fraction is less than a few percent.

In the more industrially relevant semi-dilute regime, where particle to particle interactions are important, the governing equations are less well understood.  In a influential paper \citeauthor{batchelor72d} used ensemble averaging of the bulk suspension to calculate an effective viscosity for the suspension of $\muf (1 + \nicefrac{5}{2}\phis + c \phis^2)$ with $c \approx 7.6$.  The analysis employed particle mobility functions based on the interactions of two isolated force and torque free particles, and used a `re-normalisation' technique to find the average stesslet on each sphere using the approximation that this stresslet is determined by a particle's interaction with its closest neighbour.  \citeauthor{batchelor72d} also calculated the distribution of particles surrounding a reference particle in a pure straining flow, giving important insight into the microstructure of a suspension.  \citeauthor{batchelor72d}'s techniques have been extended to include effects of microstructure anisotropy linked to surface roughness and non-hydrodynamic interaction forces by (amongst others) \citet{zinchenko84}, \citet{brady97}, \citet{wilson00}, \citet{wilson02}, \citet{wilson05} and \citet{brady06}.  While these studies provide important links between suspension microstructure and the bulk stresses existing within a suspension for specific flow fields, they have stopped short of providing frame-invariant constitutive equations that can be applied in arbitrary flow fields.  Also, as the averaging is performed over the entire suspension, these studies do not distinguish between phase velocities and hence are not able to predict particle migration.  Further, the suspension stresses predicted via these studies generally underpredict measurements, even in the semi-dilute regime, suggesting that models based on strict interpretations of force-free binary particle hydrodynamic interactions may not capture all of the relevant physics.

From a different perspective, particle migration within a range of simple flow fields and across a range of particle concentrations has been predicted by the suspension balance (SB) model, which uses phemological rheological equations to calculate averaged particle stresses, rather than stress closure models based on direct microhydrodynamic averaging \citep{nott94,morris98,morris99}.  These models use a low-inertia suspension balance equation, consistent with \citeauthor{batchelor72d}'s suspension stress equation, combined with a force balance equation for the particle phase, allowing particle migration to be predicted.  Normal viscosity coefficients fitted to experimental data have produced good agreement between experimentally measured and simulated particle migration results, at least for specific unidirectional and curvilinear flow fields.  Initial implementations of the method used the same particle stress in the particle phase equation as used in the fluid equation, however a volume averaging analysis by \citet{nott11} (with some similarity to \citeauthor{jackson97}'s analysis and dilute closure) has shown that these stresses are distinct.  The SB equations have been adapted to arbitrary two-dimensional flow fields using a technique that calculates local stress tensors within local coordinates that are aligned with the principal axes of the local flow\citep{miller09}.  This is a worthwhile approach but does involve the complexity of an eigenvector analysis at each point in space and time that stresses are required.  The SB and related diffusive flux (DF) \citep{phillips92} models are discussed in greater detail in paper II \citep{noori24a}, and compared against results generated from the present study.

In the concentrated regime, characterised by solid volume fractions higher than approximately 40\%, particles interact directly via contact and friction forces as well as hydrodynamically.  As highlighted in reviews on this research field\citep{morris20a,lemaire23}, this leads to even more complex behaviour such as shear-thinning and jamming.  In terms of frame-invariant constitutive tensorial equations, some works are based on the concept of tracking suspension structure, based on the displacement between closely interacting particle pairs \citep{phanthien95a,phanthien99,gillissen18,gillissen19a}.  Concentrated suspensions are not the focus of this work.

Reviewing this literature shows that while our understanding of suspension stresses within the semi-dilute regime has advanced considerably, a) there is uncertainty regarding the exact form of the individual phase equations that are required to predict (e.g.) particle migration, and b) there is no analytical link between particle properties and bulk suspension behaviour that is valid outside of simple flow geometries.  This limits our ability to design technologies or optimise processes that utilise these industrially relevant semi-dilute suspensions.  This paper addresses this knowledge gap: Specifically, we derive constitutive equations that can predict the velocities of the two interpenetrating phases in arbitrary flow fields by averaging closure models of microstructure processes.

In section \ref{sec:averaged_equations} we briefly describe the volume averaging technique of \citet{jackson97} and derive the multifluid fluid, solid and mixture momentum equations applicable to mono-disperse colloidal suspensions in the semi-dilute regime.  Frame-invariant closure models are then detailed that represent the stress surrounding each particle as a sum of stresses due to fluid movement through a fixed bed of particles (section \ref{sec:fixed_analysis}), and those due to interactions between particles (section \ref{sec:interaction_analysis}).  Finally, an order of magnitude analysis is performed on the derived momentum equations to determine what terms are significant (section \ref{sec:resulting_equations}).  In a companion paper (referred to as paper II) \citep{noori24a} we analyse the performance of the derived constitutive equations by calculating the resulting shear and normal stresses as a function of suspension microstructure/particle properties, and evaluate their ability to predict shear induced migration (SIM) in benchmark pressure driven flows.

\section{Averaged Equations of Motion\label{sec:averaged_equations}}

In this section we derive averaged equations of motion for each of the two phases from the local equations of motion.  The analysis applies to a suspension of solid spherical particles within an incompressible Newtonian fluid.  The particles have a radius of $a$ and are much smaller than the containing vessel.  We utilise averaging methods first detailed in \citet{anderson67} but that were more rigorously applied in \citet{jackson97} to a dilute (order $\phis$) suspension of hard spheres.  The main novelties in this section are the definition of the particle stress terms that appear in the averaged momentum equations, and the quantification of the averaging error inherent to each momentum equation.  These, along with details the averaging techniques, support subsequent analysis.

The \citet{anderson67} technique defines various averages using a local weighting function, $g(r)$.  This function is defined to be a monotonic decreasing function of $r$.  No specific form for $g(r)$ is required, however its amplitude and radius obey
\begin{equation}
4\pi \int_0^{\infty} g(r) r^2 dr = 1 \quad\text{and}\quad \int_0^l g(r) r^2 dr = \int_l^{\infty} g(r) r^2 dr
\label{eq:g_properties}
\end{equation}
Hence $l$ is a measure of the radius of the weighting function.

Several types of averages are required to derive the averaged equations.  The \emph{fluid phase average} is a volume average performed over the fluid phase within a suspension region, relating a point property $f$ at $\y$ to its average at $\x$ via
\begin{equation}
\phif(\x) \ave[f]{f}(\x) = \int_{\Vf} f(\y) g(|\x-\y|) d\scali[\y]{V}
\label{eq:fluidave_def}
\end{equation}
Here $\phif$ is the average volume fraction of the fluid phase, which can be evaluated by setting $f=1$ in this equation.   The fluid phase volume within the suspension is represented by $\Vf$.  A \emph{solid phase (volume) average}, $\ave[s]{f}$, analogous to equation (\ref{eq:fluidave_def}), is also defined with the integral taken over the solid phase volume $\Vs$.  The corresponding solid phase volume fraction is $\phis$, resulting in $\phis=1-\phif$ exactly.  Finally a \emph{mixture volume average} $\ave[m]{f}=\ave[f]{f}+\ave[s]{f}$ is defined using an integral taken over the entire suspension volume $V$.  \emph{Particle phase averages} are defined separately as ensemble averages relating a property $\fp$ associated with particle $p$ to the average property at $\x$ via
\begin{equation}
n(\x) \ave[p]{f}(\x) = \sum_p \fp g(|\x-\xp|)
\label{eq:particleave_def}
\end{equation}
Here $n$ is the average particle number density (defined by setting $\ave[p]{f}=\fp=1$ in equation (\ref{eq:particleave_def})), $\nu=4\pi a^3/3$ is the particle volume and $\Vp$ represents the volume associated with particle $p$.  The sum is taken over all particles $p$ in the domain, each with centroid $\xp$.  Equation (\ref{eq:particleave_def}) does not place any constraints on how the particle property $\fp$ is related to spatially varying local particle properties, however for velocity we require that $\up$ represents the velocity of the centroid of particle $p$, that is $\up=\vect{u}(\xp)$.  Note that in terms of notation a superscript $p$ refers to a specific particle (e.g., $\xp$, $\up$ and $\Vp$), while the subscripts of p, s, f and m, represent particle, solid, fluid and mixture phase or averages, respectively (e.g., $\ave[p]{\vect{u}}$, $\phif$, $\us$, $\um$ or $\Vs$).

There are several assumptions employed in all the averaging analyses.  Firstly, a separation of three lengthscales:  Namely $a \ll l \ll L$, where $a$ is the particle radius, $l$ is the previously defined radius of the averaging weighting function, and $L$ is a lengthscale over which averaged variables vary (ie, a lengthscale of the macroscopic flow).  Further, any analysis based on any of the averaging definitions is only valid in regions that are located far ($\gg l$) from any boundaries, such that average variable values do not depend on conditions that are outside of the suspension region.  This assumption is utilised not only in the averaging definitions, but also in derivations for identities of the averages of temporal and spatial derivatives, as given in \citet{anderson67}.  Various other averaging identities that are used in the following are derived in Appendix \ref{sec:averaging_identities}.

Turning to the averaged equations of motion:  \citet{jackson97} has already derived averaged continuity equations for each phase from the local continuity equation.  The derivation involves applying the fluid and solid averaged temporal and spatial derivative identities (equations (4) and (5) in \citet{jackson97}) to a species invariant property (ie, $f=1$) while noting the local incompressibility constraint $\vnabla \cdot \vect{u} =0$.  This produces for the fluid and solid phases,
\begin{equation}
\frac{\partial \phif}{\partial t} + \vnabla \cdot \phif \uf = 0 \quad \text{and} \quad \frac{\partial \phis}{\partial t} + \vnabla \cdot \phis \us = 0
\label{eq:continuity}
\end{equation}
respectively, where $\uf=\ave[f]{\vect{u}}$ is the fluid volume averaged velocity and $\us=\ave[s]{\vect{u}}$ is the solid volume averaged velocity.  This derivation is exact provided the region is far from boundaries.  Summing these equations leads to the mixture based incompressibility statement,
\begin{equation}
\vnabla \cdot \um = 0
\label{eq:continuitym}
\end{equation}
where $\um=\phif\uf+\phis\us$ is the mixture averaged velocity.  Note that a particle-number-based continuity equation is also derived in \citet{jackson97} (their equation (15)), however not directly used in this study.

The averaged momentum equations are similarly derived by averaging local momentum conservations over their respective phases.  Specifically, performing a fluid phase average of the local Cauchy momentum equation produces
\begin{multline}
\rhof \left [ \frac{\partial \phif \uf }{\partial t}  + \vnabla \cdot \phif \uf \uf \right ] = - \vnabla \phif \pf + \muf \vnabla \cdot \tensgammadotm \\
+ \vect[rst,f]{f} + \vect[tot,f]{f} + \rhof \phif \g + \epsf
\label{eq:momentumf1}
\end{multline}
where $\rhof$ and $\muf$ are the density and viscosity of the fluid phase, respectively, $\pf=\ave[f]{p}$ is the fluid volume averaged pressure, $\tensgammadotm = \vnabla\um + (\vnabla\um)^T$ is the symmetric, trace-less (due to equation (\ref{eq:continuitym})) strain rate based on mixture velocities, and $\g$ is the acceleration due to gravity.  An error $\epsf$ of $\order{\phif \rhof l \uf^2/L^2}$ in this equation has resulted solely from averaging the inertial advection term, as derived in Appendix \ref{sec:averaging_identities} and summarised by equation (\ref{eq:aveiduuf}).  Also, the stress within a Newtonian fluid $\tens{\sigma}=-p\Itens+\muf (\vnabla\vect{u} + (\vnabla\vect{u})^T)$ has been applied within the fluid phase to produce the averaged pressure and viscous stress terms\citep{jackson97}.  As detailed in \citet{joseph90}, the form of these terms is consistent with those derived using alternative ensemble averaging approaches (see also \citet{drew98}).


Two forces are defined in equation (\ref{eq:momentumf1}).  The first of these,
\begin{equation}
\vect[rst,f]{f} = - \rhof \vnabla \cdot \phif \ave[f]{\uf' \uf'}
\label{eq:frstf_def}
\end{equation}
is a `Reynolds' type stress term that results from decomposing and averaging the inertial advection term.  Here $\uf'=\vect{u}-\uf$ is the difference between the local and average fluid velocities, defined only within the fluid.
The second force represents the effect of particle surface stresses on the fluid phase, and is given by
\begin{equation}
\vect[tot,f]{f} = - \sum_p \intSp \vect[tot]{t}^p (\y) g (|\x-\y|) d \scali[\y]{S}
\label{eq:ftotf_def}
\end{equation}
where $\vect[tot]{t}^p(\y) = \np(\y) \cdot \tens{\sigma} (\y)$ is the total surface traction acting at $\y$ on the particle's surface, $\np$ is an unit normal pointing outwards from the particle%
, and $S^p$ is the surface area, all associated with particle $p$.  While we focus on particle surface stresses that originate from hydrodynamic interactions in this study, we note that equation (\ref{eq:ftotf_def}) applies regardless of the origin of $\tens{\sigma}$.

Forming a solid phase momentum equation is slightly more involved.  Adapting the method of \citet{anderson67}, particle phase averaging of Newton's second law applied to a particle $p$ gives
\begin{equation}
\rhos \nu \left [ \frac{\partial n \ave[p]{\vect{u}} }{\partial t}  + \vnabla \cdot n \ave[p]{\vect{u} \vect{u}} \right ] = \vect[tot,s]{f} + \rhos n \nu \g
\label{eq:momentums1}
\end{equation}
where the average force exerted by the fluid on the solid phase is given by 
\begin{equation}
\vect[tot,s]{f} = \sum_p g(|\x-\xp|) \intSp \vect[tot]{t}^p(\y) d \scali[\y]{S}
\label{eq:ftots_def}
\end{equation}
While equation (\ref{eq:momentums1}) is exact, we wish to express the momentum balance in terms of solid phase, rather than particle phase averaged quantities, and doing these conversions introduces averaging errors.  Specifically, using equivalences derived in Appendix \ref{sec:averaging_identities} for $\n \nu \approx \phis$, $\ave[p]{\vect{u}} \approx \us$ and $\ave[p]{\vect{u}\vect{u}} \approx \us\us + \ave[s]{\us' \us'}$, given by equations (\ref{eq:aveidphis}), (\ref{eq:aveidus}) and (\ref{eq:aveiduup}), respectively, yields
%
%
\begin{equation}
\rhos \left [ \frac{\partial \phis \us }{\partial t}  + \vnabla \cdot \phis \us \us \right ] = \vect[rst,s]{f} + \vect[tot,s]{f} + \rhos \phis \g + \epss
\label{eq:momentums2a}
\end{equation}
where another `Reynolds' type stress force
\begin{equation}
\vect[rst,s]{f} = - \rhos \vnabla \cdot \phis \ave[s]{\us' \us'}
\label{eq:frsts_def}
\end{equation}
is defined in terms of the solid perturbation velocity $\us'=\up-\us$, $\up$ is the local velocity of the relevant particle $p$, and the error $\epss = \order{\phis \rhos a \us^2/L^2,\phis \rhos a^2 \g/L^2}$ has resulted from averaging the inertial and weight terms (noting $\order{a,b,...}=\order{a}+\order{b}+...$).


In addition to the fluid and solid phase momentum equations, equations (\ref{eq:momentumf1}) and (\ref{eq:momentums2a}) can be added to yield a mixture momentum equation,
%
%
\begin{multline}
\frac{\partial \rhom \vecti[\rho]{u} }{\partial t}  + \vnabla \cdot \rhom \vecti[\rho]{u} \vecti[\rho]{u} = - \vnabla \phif \pf + \muf \vnabla \cdot \tensgammadotm + \vecti[\rho,\text{m}]{f} + \vect[rst,f]{f} + \vect[rst,s]{f} \\
 + \vect[tot,f]{f} + \vect[tot,s]{f} + \rhom \g + \epsf + \epss
\label{eq:momentumm1}
\end{multline}
where
\begin{equation}
\vecti[\rho,\text{m}]{f} = - \vnabla \cdot \frac{\rhof \rhos \phif \phis}{\scal[m]{\rho}} ( \us - \uf )( \us - \uf )
\label{eq:frhom_def}
\end{equation}
is an inertial correction term generated by summing the phase averaged momentum advection terms\citep{jackson97}
and $\vecti[\rho]{u}=(\rhof \phif \uf + \rhos \phis \us)/\scal[m]{\rho}$ is the mass averaged suspension velocity.

\begin{figure}
\centering
\def\svgwidth{0.65\textwidth}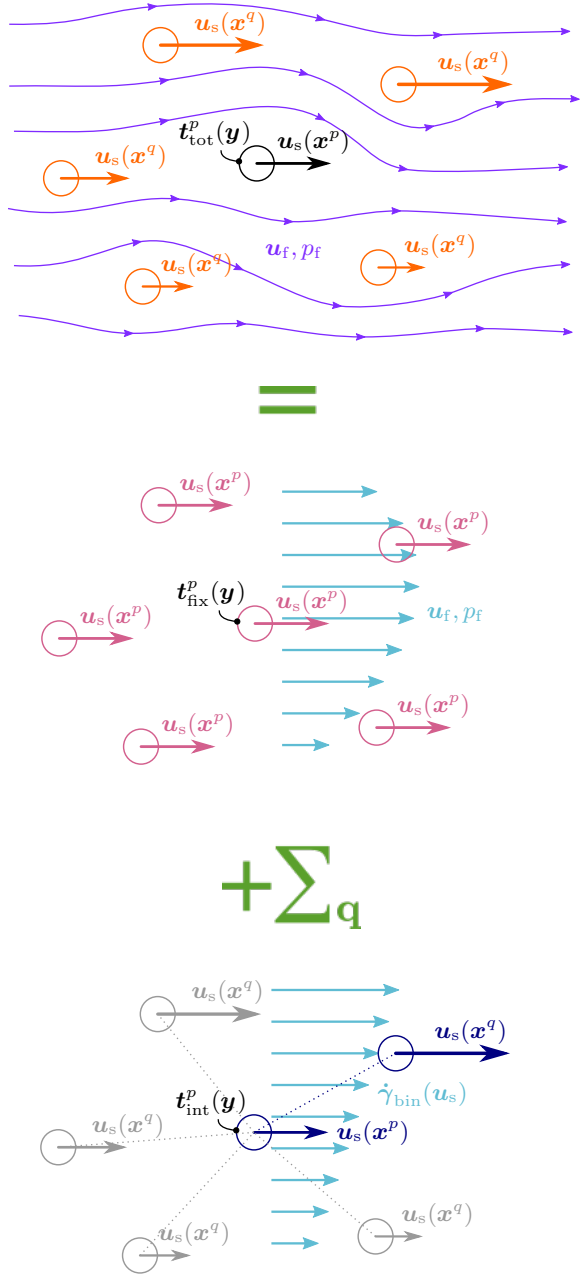
\caption{The total traction ($\vect[tot]{t}^p$) acting on the surface (at $\y$) of a reference particle $p$ centred at $\xp$ and surrounded by secondary particles $q$ each centred at $\xq$ is decomposed into that resulting from a fixed array of surrounding particles ($\vect[fix]{t}^p$) and that resulting from the relative movement of the surrounding particles ($\vect[int]{t}^p$) calculated by summing binary interactions.}
\label{fig:stress_decomposition}
\end{figure}

To proceed further we make two key assumptions regarding the stress experienced by each particle.  Working directly in particle surface tractions, we firstly assume that the total stress acting on the surface of the reference particle can be expressed as a function of the averaged fluid and solid fields, $\uf$, $\us$, $\phis$ and $\pf$.  Secondly, as illustrated in the schematic Figure \ref{fig:stress_decomposition}, we then partition this traction into two terms,
\begin{equation}
\vect[tot]{t}^p(\y,\uf,\us,\phis,\pf)= \vect[fix]{t}^p(\y,\uf,\us(\xp),\pf,\phis) + \vect[int]{t}^p(\y,\uf,\us,\pf,\phis)
\label{eq:sigma_decomposition}
\end{equation}
where $\vect[fix]{t}^p$ represents the traction exerted on the reference particle due to (confined) fluid movement, and $\vect[int]{t}^p$ represents the traction exerted on the reference particle due to relative particle movement.  More rigorously, the `fixed' traction $\vect[fix]{t}^p$ is defined as that on the surface of particle $p$ (centred at $\xp$) when the particle is travelling at velocity $\us(\xp)$ in the presence of all other particles, but with the other particles all travelling at the reference particle velocity of $\us(\xp)$, and all within a fluid having the average velocity and pressure fields of $\uf$ and $\pf$, respectively.  In contrast, the `interaction' traction $\vect[int]{t}^p$ is defined as the traction experienced by a particle $p$ when all other particles (i.e. $q$ at $\xq$) are moving on average at their local $\us(\xq)$ velocity, minus the traction experienced by particle $p$ in the same situation but where all other particles are moving at the velocity of the reference particle, $\us(\up)$.  In both situations all particles again exist within a fluid having the average velocity and pressure fields of $\uf$ and $\pf$, respectively.

Aside from the dependence on average field variables, the traction decomposition represented by equation (\ref{eq:sigma_decomposition}) is exact, however, approximations are introduced later when closure models are developed to describe the two component tractions.  Specifically the forces resulting from $\vect[fix]{t}$ are evaluated in section \ref{sec:fixed_analysis} using known results for a single particle moving in a quadratic velocity field, but modified using a closure model that accounts for the effects of the fixed surrounding particles on the form and skin drag acting on the reference particle.  Similarly, the forces resulting from $\vect[int]{t}$ are evaluated in section \ref{sec:interaction_analysis} by summing the contributions from binary particle interactions occurring in a linear fluid velocity field, with this field so chosen to ensure that the relative movement between interacting particles is consistent with the solid velocity field $\us$.  The velocity fields shown in figure \ref{fig:stress_decomposition} represent those used in these closure models.

Before turning to these stress evaluations, however, we derive relationships between the fluid and solid particle forces that will aid subsequent analysis.  We first define particle forces that are analogous to equations (\ref{eq:ftotf_def}) and (\ref{eq:ftots_def}) but are decomposed using equation (\ref{eq:sigma_decomposition}) into separate components as
\begin{equation}
\vecti[j,\text{f}]{f} = - \sum_p \intSp \vecti[j]{t}^p (\y) g (|\x-\y|) d \scali[\y]{S}
\label{eq:fjf_def}
\end{equation}
and
\begin{equation}
\vecti[j,\text{s}]{f} = \sum_p g(|\x-\xp|) \intSp \vecti[j]{t}^p (\y) d \scali[\y]{S}
\label{eq:fjs_def}
\end{equation}
where $j$ = fix or int.  Noting equation (\ref{eq:sigma_decomposition}), this gives
\begin{equation}
\vecti[\text{tot},i]{f} = \vecti[\text{fix},i]{f} + \vecti[\text{int},i]{f} \quad \text{for $i$ = s or f}
\label{eq:totf}
\end{equation}
Proceeding in a similar vein to \citet{jackson97} by expanding $g (|\x-\y|)$ using a taylor series about the centre of the reference particle $\xp$, and utilising the results $\vnablaxp^n g(|\x-\xp|) = (-1)^n \vnablax^n g(|\x-\xp|)$ and $(\np(\y) \cdot \vnablax)^m g(|\x-\xp|) = (\vnablax \cdot)^m (g(|\x-\xp|) \np(\y)^m)$, equation (\ref{eq:fjf_def}) can be written
\begin{equation}
\vecti[j,\text{f}]{f} = - \sum^{\infty}_{m=0} \frac{(-1)^m}{m!} (\vnablax \cdot)^m \sum_p g(|\x-\xp|) a^m \intSp {(\np(\y))}^m \vecti[j]{t}^p (\y) d \scali[\y]{S}
\label{eq:fjf_parts1}
\end{equation}
Recognising that the first term in this series is $\vecti[j,\text{s}]{f}$, mixture based particle forces can be defined via
\begin{equation}
\vecti[j,\text{m}]{f} = \vecti[j,\text{f}]{f} + \vecti[j,\text{s}]{f} = \vnabla \cdot \tensi[j,\text{s}]{f}' - \frac{1}{2} \vnabla \cdot \left ( \vnabla \cdot \tensi[j,\text{s}]{f}'' \right ) + \order{\frac{a^3}{L^3} \vecti[j,\text{s}]{f}}
\label{eq:fjf_parts}
\end{equation}
where a more general solid particle force (moment) is defined by
\begin{equation}
\vecti[j,\text{s}]{f}^{'^m} = a^m \sum_p g(|\x-\xp|) \intSp {(\np(\y))}^m \vecti[j]{t}^p (\y) d \scali[\y]{S}
\label{eq:fjs'm_def}
\end{equation}
that applies for $m \ge 0$ and $j$ = fix or int.  As shown by equation (\ref{eq:fjf_parts}), evaluation of the mixture, fluid and solid particle fixed and interaction forces requires evaluation of the corresponding $\vecti[j,\text{s}]{f}^{'^m}$, as given by equation (\ref{eq:fjs'm_def}), for $m=0,1$ and $2$.  This is the objective of the next two sections.


\section{Fixed particle forces\label{sec:fixed_analysis}}

\subsection{Fixed particle stress model closure assumptions\label{sec:fixed_assumptions}}

As mentioned above, integrals of the traction around the reference particle when surrounded by a fixed array of particles, required by equation (\ref{eq:fjs'm_def}) (with $j=$ fix), are calculated based on the motion of a force-free sphere in a quadratic velocity field, but modified to account for the presence of the surrounding particles.  \Citet{nadim91} provides a solution for the traction surrounding a single force-free sphere moving within a quadratic velocity field.  \Citet{jackson97} has already added to this linear velocity field and weight effects\citep{leal92}, and incorporated it into the suspension equations to model dilute (that is, non-interacting) particle stresses.  However there are some difficulties with this approach that resulted in a correction to the original paper\citep{jackson98}, and inconsistencies that remain when the suspension is not dilute that we wish to address.

Importantly we make three modifications (closure assumptions) to \citeauthor{jackson97}'s analysis to account for the presence of the surrounding particles on the traction integrals.

The discussed traction solution is based on the Stokes equations surrounding a single sphere.  Our first modification recognises that the far-field pressure field local to the reference particle, which determines the form drag acting on the particle, is not consistent with the pressure field existing within the suspension at the particle's location.  There are two reasons why the dilute local and suspension pressure fields are inconsistent (as employed in \citet{jackson97}, for example).  Firstly by basing the local traction calculation on the Stokes equations, pressure gradients due to acceleration (changes in inertia) are not captured at a local level, in contrast to the suspension equations which do account for inertia and hence generate pressure gradients due to acceleration.  Indeed, accounting for acceleration at the local level was the purpose of the \citet{jackson98} correction to the original dilute suspension derivation.  Secondly, in the presence of other particles the local pressure gradient is further modified from the pressure gradient acting on the reference particle in isolation, due to (e.g.) drag forces from the other particles acting on the fluid flow.  This effect is not included in either of the \citeauthor{jackson98} derivations.  Below we account for both these inertial and surrounding particles' effects on the local pressure gradient by introducing a body force $\bp$ into the locally solved Stokes equations, with the force chosen so that the local pressure gradient used in the quadratic field traction calculation matches the fluid phase suspension pressure gradient, evaluated at the location of the reference particle $p$.  This modification ensures that whatever forces and (potentially) additional physics are introduced to the suspension scale momentum equations, the resulting pressure gradient will be faithfully reproduced on the local particle scale.

The second modification to the dilute traction solution recognises the effect that the surrounding fixed particles have on the skin drag experienced by the reference particle.  Specifically, for a given fluid phase velocity field $\uf$ that surrounds the reference particle, the effect of a fixed array of surrounding particles is to decrease the volume available through which the surrounding fluid can flow, which in turn increases the fluid strain rates existing on the surface of the reference particle.  Viscous stresses experienced by the reference particle are the product of fluid viscosity $\muf$ and local strain rates.  Hence, to capture the increase in strain rates on the reference particle due to the fixed array of particles in our closure model, we multiply the fluid viscosity in the traction expression by a factor $\scal[fix]{\lambda}$ which increases with the local particle concentration.  That is, within the dilute traction expression we use the local viscosity
\begin{equation}
\mu(\y) =\scal[fix]{\lambda}(\phis(\y)) \muf
\label{eq:lambda_fix}
\end{equation}
rather than the fluid viscosity $\muf$.  In paper II \citep{noori24a} we show how $\scal[fix]{\lambda}$ can be related to measured suspension properties such as the hindered settling function or mixture shear viscosity.  Note that this closure model uses a modified viscosity to capture the increase in particle traction due to increased local strain rates, rather than representing the traction on a sphere in a variable viscosity fluid.

The final modification to the local traction expression is concerned with how the fluid phase velocity field, $\uf$, is used to represent the local fluid velocity field surrounding the reference particle, given that the fluid phase velocity field $\uf$ is not divergence free.  The \citet{nadim91} and \citet{leal92} analyses on which the traction expression is based utilise divergence-free properties of the local field velocity characterisation tensors, so the $\uf$ field cannot be used directly with the traction expression.  As a closure assumption we propose physically reasonable adjustments to the suspension scale $\uf$ field to create local velocity field vectors and tensors that are consistent with a divergence free velocity field, but can be used in the reference particle traction expression.


\subsection{Evaluating the fixed particle traction\label{sec:fixed_traction}}


The following analysis is adapted from \citet{jackson97}'s single sphere solution which incorporates both linear and quadratic velocity field terms\citep{nadim91,leal92}.  With the introduction of a general body force the equations that are solved local to reference particle $p$ are
\begin{equation}
\vnabla \cdot \tens{\sigma} + \bp = \vect{0}, \text{ where }\tens{\sigma}=-p\Itens + \mu \left [ \vnabla\vect{u} + \left ( \vnabla\vect{u} \right )^T \right ]
\label{eq:stokesclosure}
\end{equation}
and as discussed above $\bp$ is a body force acting on the fluid that accounts for gravitational, inertial and potentially non-dilute form drag effects.  Velocity boundary conditions local to the particle are specified by expanding the local undisturbed velocity $\vect{u}^\infty$ via a multipole expansion, which adopting a modified \citet{nadim91} notation is
\begin{equation}
\vect{u}^\infty(\rfix) = \Up + \Omegap \cross \rfix + \rfix \cdot \Ep + \rfix\rfix : \Kp + \order{\rfix\rfix\rfix\vnabla\vnabla\vnabla \vect{u}^\infty}
\label{eq:farfieldv}
\end{equation}
where $\rfix$ is the position relative to the particle centre, $\Up=\vect{u}^\infty(\rfix=\vect{0})$ is the undisturbed velocity at the particle centre, and
\begin{equation}
\Omegap = \frac{1}{2} \left [ \vnabla \cross \vect{u}^\infty \right ], \Ep = \frac{1}{2} \left [ \vnabla \vect{u}^\infty + (\vnabla \vect{u}^\infty)^T \right ] \text{ and }
\Kp = \frac{1}{2} \left [ \vnabla \vnabla \vect{u}^\infty \right ]
\label{eq:closureveldefs}
\end{equation}
are functions of the undisturbed velocity field also evaluated at the particle centre.  The traction around the sphere is expressed in terms of these velocity field vectors and tensors\citep{jackson97,nadim91}.  For consistency with the solution around the force-free sphere, the pressure field in the absence of the sphere must also obey the modified Stokes equations.  Hence the far-field pressure field around reference particle $p$ is given by
\begin{equation}
p^\infty(\rfix) = \Pp + \bp \cdot \rfix + 2\mu\Itens:\Kp\cdot\rfix + \order{\mu\rfix\rfix\vnabla\vnabla\vnabla \vect{u}^\infty}
\label{eq:farfieldp}
\end{equation}
where $\Pp=p^\infty(\rfix=\vect{0})$ is the undisturbed pressure at the particle centre location.  We note that by employing the Stokes equations, this closure model neglects the influence of inertia on the local particle scale, so that inertial lift forces (for example) are not accounted for in the final suspension averaged equations.  Also, as is assumed throughout the entire averaging process, particles are assumed to be far from any system boundaries and hence wall induced particle forces are also not accounted for in this analysis.  With this local problem definition the traction surrounding reference particle $p$ at location $\vect{y} = a \np + \xp$ can be adapted from the previous works as
\begin{align}
\tfixp ( \vect{y} ) & = \frac{3\mu}{2a}(\Up-\up) + 3\mu ( \Itens + \np\np ) \cdot \left [ ( \Omegap - \omegap ) \cross \np \right ] \nonumber \\
& + 5 \mu \np \cdot \Ep - \np \Pp - a \np\np \cdot \bp \nonumber \\
& + \mu a \left [ \frac{35}{4} \np\np : \gammap + \frac{5}{3} \np \cross (\np \cdot \thetap) - 3 \np\np \cdot \taup + \frac{3}{2}\taup \right ] \nonumber \\
& + \order{ \mu a^2 \np\np\np \vnabla\vnabla\vnabla \vect{u}^\infty }
\label{eq:traction1}
\end{align}
where the quadratic field vector and tensors are defined by \citep{nadim91}
\begin{align}
\gamma_{ijk}^p & = \frac{1}{3}( K_{ijk}^p + K_{kij}^p + K_{jki}^p ) - \frac{1}{15}( \delta_{ij}K_{ppk}^p + \delta_{ik}K_{ppj}^p + \delta_{jk}K_{ppi}^p ) \label{eq:tractiongamma} \\
\theta_{lm}^p & = K_{lpq}^p \epsilon_{qpm} + K_{mpq}^p \epsilon_{qpl} \label{eq:tractiontheta} \\
\tau_l^p & = K_{ppl}^p
\label{eq:tractiontau}
\end{align}
Closure involves relating the local particle scale properties $\Pp$, $\bp$, $\Up$, $\Omegap$, $\Ep$ and $\Kp$, as well as the local viscosity $\mu$, to suspension scale variables.

Dealing first with the body force $\bp$, differentiation of equation (\ref{eq:farfieldp}) gives
\begin{equation}
\vnabla p^\infty = \bp + 2\mu\taup + \order{\mu\rfix\vnabla\vnabla\vnabla \vect{u}^\infty}
\label{eq:farfielddp}
\end{equation}
Employing the closure assumption that the local far field pressure gradient and values must be consistent with the suspension scale pressure gradient and values yields
\begin{equation}
\bp = \vnabla \pf - 2\mu\taup + \order{\mu\rfix\vnabla\vnabla\vnabla \vect{u}^\infty} \text{ and } P^p = \pf
\label{eq:farfielddp2}
\end{equation}
where in the following it is understood that both $\vnabla \pf$ and $\pf$ are evaluated at $\xp$.

Turning next to the undisturbed velocity vectors and tensors, those that do not depend on the trace of the strain rate are equated to the local fluid phase values evaluated at $\xp$, giving
\begin{equation}
\Up = \uf \text{ and } \Omegap = \omegaf = \frac{1}{2} \vnabla \cross \uf
\label{eq:farfieldv1}
\end{equation}
where $\omegaf$ is the suspension scale fluid rotation vector.  The two remaining tensors require certain properties for the traction analysis to be valid\citep{nadim91}, specified as
\begin{equation}
E^p_{ii} = 0, K^p_{ipp} = 0 \text{ and } K^p_{ijk}=K^p_{jik}
\label{eq:farfieldprops}
\end{equation}
For $\Ep$ we remove a multiple of the trace of the strain rate $\vnabla \cdot \uf$ from the diagonal elements of the fluid phase strain rate tensor to satisfy equation (\ref{eq:farfieldprops}), yielding
\begin{equation}
\Ep = \frac{1}{2} \left [ \tensgammadotf \right ]_\text{D} \label{eq:farfieldEp}
\end{equation}
where
\begin{equation}
\tensgammadotf = \vnabla \uf + ( \vnabla \uf )^T - \gammadotsphf \ItensD \quad \text{ and } \quad \gammadotsphf = \frac{2}{N} \vnabla \cdot \uf
\label{eq:tensgammaf_def}
\end{equation}
Here $N$ is the number of physical dimensions considered in the suspension scale problem and $\ItensD = \left [ \Itens \right ]_\text{D}$ is the unit tensor in the physical dimensions.  In the above we have introduced a new notation, namely $[...]_\text{D}$, which means to evaluate all terms included in the brackets in the physical dimensions of the suspension scale problem, leaving any terms related to other dimensions zero.  This notation is used to convert suspension scale velocity fields, which may be two dimensional (e.g., flow through an infinite slit), to consistent three dimensional velocity fields which are used at the particle scale to calculate the fixed particle forces.  Specifically, equation (\ref{eq:tensgammaf_def}) ensures that any correction to the local flow field around the particle due to divergence of the suspension scale velocity field is confined to the physical dimensions of the problem.  The physical dimension notation is detailed more fully in Appendix \ref{sec:dimensional_operator}.

The correction for $\Kp$ is slightly more complex than $\Ep$ due to the combination of symmetry and divergence free conditions.  Restricting the form of the correction to first order derivatives of $\vnabla \cdot \uf = N \gammadotsphf /2$ a suitable definition that satisfies equation (\ref{eq:farfieldprops}) is
\begin{equation}
\Kp = \frac{1}{2} \left [ \vnabla \vnabla \uf - \frac{N}{2(2+N)} \left ( \delta_{ij}\frac{\delta \gammadotsphf}{\delta x_k} + \delta_{ik}\frac{\delta \gammadotsphf}{\delta x_j} + \delta_{jk}\frac{\delta \gammadotsphf}{\delta x_i} \right )\vecti[i]{\delta}\vecti[j]{\delta}\vecti[k]{\delta} \right ]_\text{D}
\label{eq:farfieldKp}
\end{equation}
Again, fluid velocity gradients required in equations (\ref{eq:farfieldEp}) and (\ref{eq:farfieldKp}) are evaluated at $\xp$ and in the physical dimensions of the system.  Note that the definitions for both $\Ep$ and $\Kp$ ensure that for divergence free fluid phase velocity fields, the particle scale far-field velocity field reduces to the suspension scale fluid velocity field.

Finally, the effective viscosity $\mu$ used in equation (\ref{eq:traction1}) is assumed to vary around the surface of the reference particle according to the closure model discussed in Section \ref{sec:fixed_assumptions}.  This is achieved by expanding equation (\ref{eq:lambda_fix}) around $\xp$ using a Taylor series, giving for $\vect{y} = a \np + \xp$
\begin{align}
\mu (\vect{y}) &= \muf \lambdafixp + \muf a \np \cdot \vnabla \lambdafixp + \muf a^2 \frac{1}{2} \np\np : \vnabla\vnabla \lambdafixp \nonumber \\
& + \order{ \muf a^3 \np\np\np\vnabla\vnabla\vnabla \lambdafixp}
\label{eq:lambda_fix2}
\end{align}
Here $\lambdafixp$ and derivatives thereof are functions of the averaged variable $\phis$ (according to equation (\ref{eq:lambda_fix})) and are evaluated at $\xp$.

The traction acting on the surface of reference particle $p$ can now be fully specified in terms of suspension scale variables by substituting equations (\ref{eq:farfielddp2}), (\ref{eq:farfieldv1}), (\ref{eq:farfieldEp}), (\ref{eq:farfieldKp}) and (\ref{eq:lambda_fix2}) into equation (\ref{eq:traction1}), yielding
\begin{align}
\tfixp ( \vect{y} ) & = \frac{3}{2a}\muf\lambdafixp(\uf-\up) + \frac{3}{2}\muf(\uf-\up) \np \cdot \vnabla\lambdafixp \nonumber \\
& + \frac{3}{4}\muf a(\uf-\up) \np\np : \vnabla\vnabla\lambdafixp \nonumber \\
& + \order{\muf a^2 (\uf-\up) \np\np\np \vnabla\vnabla\vnabla\lambdafixp } \nonumber \\
& + 3\muf \lambdafixp ( \Itens + \np\np ) \cdot \left [ ( \omegaf - \omegap ) \cross \np \right ] \nonumber \\
& + 3\muf a \np \cdot \vnabla \lambdafixp ( \Itens + \np\np ) \cdot \left [ ( \omegaf - \omegap ) \cross \np \right ] \nonumber \\
& + \order{ \muf a^2 \np\np\np \vnabla\vnabla \lambdafixp ( \omegaf - \omegap ) } \nonumber \\
& + \frac{5}{2} \muf \lambdafixp \np \cdot \tensgammadotf + \frac{5}{2} \muf a \np \cdot \vnabla \lambdafixp \np \cdot \tensgammadotf \nonumber \\
& + \order{\muf a^2 \np\np\np \vnabla \uf \vnabla\vnabla \lambdafixp} \nonumber \\
& + \muf \lambdafixp a \left [ \frac{35}{4} \np\np : \gammap + \frac{5}{3} \np \cross (\np \cdot \thetap) + \left ( \frac{3}{2} \Itens - \np\np \right ) \cdot \taup \right ] \nonumber \\
& + \order{\muf a^2 \np\np\np \vnabla\vnabla \uf \vnabla \lambdafixp} \nonumber \\
& - \np \pf - a \np\np \cdot \vnabla \pf \nonumber \\
& + \order{ \muf a^2 \lambdafixp \np\np\np \vnabla\vnabla\vnabla \uf }
\label{eq:traction2}
\end{align}
where $\gammap$, $\thetap$ and $\taup$ are defined as per equations (\ref{eq:tractiongamma}-\ref{eq:tractiontau}) and $\Kp$ is defined via equation (\ref{eq:farfieldKp}).  In the above terms containing $a^3$ and above are not represented.

\subsection{Evaluating the fixed particle forces\label{sec:fixed_forces}}

The fixed particle forces are now evaluated by substituting $\tfixp$ from equation (\ref{eq:traction2}) into equation (\ref{eq:fjs'm_def}).

Starting with equation (\ref{eq:fjs'm_def}) with $m=0$ for $\vect[fix,s]{f}$ (or equation (\ref{eq:fjs_def})), as $\tfixp$ is integrated over the surface of particle $p$, terms within (\ref{eq:traction2}) that involve odd numbers of unit normals $\np$ do not contribute to the integral.  Further, the third term of equation (\ref{eq:traction2}) involves the same slip velocity as the first term, but is $\order{a^2/L^2}$ smaller (given that $\lambdafixp$ is a function of the averaged variable $\phis$), so can also be neglected from the integral.  This leaves five terms within $\vect[fix,s]{f}$, being
\begin{align}
\vect[fix,s]{f} & = \sum_p g(|\x-\xp|) \biggl\{ \intSp \frac{3}{2a}\muf\lambdafixp(\uf-\up) d \scali[\y]{S} \nonumber \\
& + \intSp 3\muf a \np \cdot \vnabla \lambdafixp ( \Itens + \np\np ) \cdot \left [ ( \omegaf - \omegap ) \cross \np \right ]  d \scali[\y]{S} \nonumber \\
& + \intSp \frac{5}{2} \muf a \np \cdot \vnabla \lambdafixp \np \cdot \tensgammadotf d \scali[\y]{S} \nonumber \\
& + \intSp \muf \lambdafixp a \left [ \frac{35}{4} \np\np : \gammap + \frac{5}{3} \np \cross (\np \cdot \thetap) + \left ( \frac{3}{2} \Itens - \np\np \right ) \cdot \taup \right ] d \scali[\y]{S} \nonumber \\
& - \intSp a \np\np \cdot \vnabla \pf d \scali[\y]{S} \biggr\}
\label{eq:ffixs_def2}
\end{align}
For the forth term on the right (that originated from the $\Kp$ related terms) careful algebra utilising the dimensional operators defined in Appendix \ref{sec:dimensional_operator} and the identity equations (\ref{eq:I5}), (\ref{eq:I6}) and (\ref{eq:intrel1}) leads to
\begin{align}
\intSp \np\np : \gammap d \scali[\y]{S} & = \frac{4 \pi a^2}{3} \gamma_{aal}^p \vecti[l]{\delta} = \vect{0} \nonumber \\
\intSp \np \cross (\np \cdot \thetap) d \scali[\y]{S} & = \frac{4 \pi a^2}{3} \theta_{ik}^p \epsilon_{ikl} \vecti[l]{\delta} = \vect{0} \nonumber \\
\intSp \left ( \frac{3}{2} \Itens - \np\np \right ) \cdot \taup d \scali[\y]{S} & = \frac{4 \pi a^2}{3} \frac{7}{2} \taup = \frac{4 \pi a^2}{3} \frac{7}{4} \left [ \tensgammadotf - (N-1) \ItensD \gammadotsphf \right ]
\label{eq:Kp_identities}
\end{align}
Substituting these identities, along with those from equations (\ref{eq:I2}), (\ref{eq:I6}), (\ref{eq:I9}) and (\ref{eq:I10}) produces
\begin{multline}
\vect[fix,s]{f} = \nu \sum_p g(|\x-\xp|) \biggl\{ \frac{9}{2a^2}\muf\lambdafixp(\uf-\up) + 3\muf ( \omegaf - \omegap ) \cross \vnabla \lambdafixp \\ + \frac{5}{2} \muf \vnabla \lambdafixp \cdot \tensgammadotf + \frac{7}{4} \muf\lambdafixp \vnabla \cdot \left [ \tensgammadotf - (N-1) \ItensD \gammadotsphf \right ] - \vnabla \pf \biggr\}
\label{eq:ffixs_def3}
\end{multline}
Now variables such as $\uf$, $\up$, $\omegaf$, $\omegap$, $\lambdafixp$, $\vnabla \lambdafixp$, $\tensgammadotf$ etc are all associated with particle $p$, being either properties of particle $p$ or defined in the above to be evaluated at $\xp$.  So these variables become particle averaged when combined with the leading $\sum_p g(|\x-\xp|)$ sum.  Further, using $n \nu = \phis$ from equation (\ref{eq:aveidphis}) (neglecting an $\order{a^2/L^2}$ error), $\ave[p]{\vect{u}} = \us$ from equation (\ref{eq:aveidus}) (neglecting an $\order{a/L}$ error) and utilising the repeated averaging equations (\ref{eq:aveidf1f2_2}), (\ref{eq:aveidaf1}) and (\ref{eq:aveidaf1af2}) (neglecting multiple $\order{l/L}$ errors) leads to
\begin{equation}
\vect[fix,s]{f} = - \phis \vnabla \pf + \vect[drag,s]{f} + \vect[faxen,s]{f} + \vect[rot,s]{f}
\label{eq:ffixs_def4}
\end{equation}
where
%
\begin{equation}
\vect[rot,s]{f} = 3 \phis \muf ( \omegaf - \omegas ) \cross \vnabla \lambdafix \label{eq:frots}
\end{equation}
and $\vect[drag,s]{f}$ and $\vect[faxen,s]{f}$ are defined by equations (\ref{eq:fdrags}) and (\ref{eq:ffaxens}), respectively.  This completes the derivation of $\vect[fix,s]{f}$.

The derivation of $\vect[fix,s]{f}'$ and $\vect[fix,s]{f}''$ proceeds analogously.  For $\vect[fix,s]{f}'$, after substituting the fixed particle traction from equation (\ref{eq:traction1}) into equation (\ref{eq:fjs'm_def}) with $m=1$, many terms involve an odd number of $\np$'s, and hence do not contribute to the spherical integral.  Further, the seventh term from equation (\ref{eq:traction1}) when used in the $\vect[fix,s]{f}'$ evaluation is $\order{a^2/L^2}$ smaller than the fifth term from equation (\ref{eq:traction1}), and can hence be neglected.  Similarly, the tenth, twelfth and thirteenth terms in equation (\ref{eq:traction1}) all make $\order{a^2/L^2}$ smaller contributions to $\vect[fix,s]{f}'$ than the eighth term of equation (\ref{eq:traction1}), and so can also be neglected.  After these simplifications we get
\begin{align}
\vect[fix,s]{f}' & = a \sum_p g(|\x-\xp|) \biggl\{ \intSp \frac{3}{2} \muf \np (\uf-\up) \vnabla \lambdafixp \cdot \np d \scali[\y]{S} \nonumber \\
& + \intSp 3\muf \lambdafixp \np ( \Itens + \np\np ) \cdot \left [ ( \omegaf - \omegap ) \cross \np \right ]  d \scali[\y]{S} \nonumber \\
& + \intSp \frac{5}{2} \muf \lambdafixp \np \np \cdot \tensgammadotf d \scali[\y]{S} - \intSp \np\np \pf d \scali[\y]{S} \biggr\}
\label{eq:ffixs'_def2}
\end{align}
Employing the spherical integral identity equations (\ref{eq:I5}), (\ref{eq:I9}) and (\ref{eq:I10}), and again employing the same averaging approximations as employed for $\vect[fix,s]{f}$ leads to
\begin{align}
\vect[fix,s]{f}' & = \frac{3}{2} \phis \muf \vnabla \lambdafix (\uf-\us) + 3 \phis \muf \lambdafix ( \scali[\text{f},l]{\omega} - \scali[\text{s},l]{\omega} ) \epsilon_{lik} \vecti[i]{\delta}\vecti[k]{\delta} \nonumber \\
& + \frac{5}{2} \phis \muf \lambdafix \tensgammadotf - \phis \pf \Itens
\label{eq:ffixs'_def3}
\end{align}
For $\vect[fix,s]{f}''$, after substituting equation (\ref{eq:traction1}) into equation (\ref{eq:fjs'm_def}) with $m=2$, any terms containing an even number of $\np$'s can be neglected from the traction equation (\ref{eq:traction1}).  Also, noting from equation (\ref{eq:fjf_parts}) that $\vnabla \cdot \vect[fix,s]{f}'$ and $\vnabla \cdot ( \vnabla \cdot \vect[fix,s]{f}'' )$ will be combined to form $\vect[fix,m]{f}$, we find that all remaining terms from equation (\ref{eq:traction1}) used in the calculation of $\vect[fix,s]{f}''$ have counterparts in $\vect[fix,s]{f}'$ that are $\order{L^2/a^2}$ larger, except for the first term.  Hence the calculation for $\vect[fix,s]{f}''$ simplifies to
\begin{equation}
\vect[fix,s]{f}'' = a \sum_p g(|\x-\xp|) \intSp \frac{3}{2} \muf  \lambdafixp \np \np (\uf-\up) d \scali[\y]{S}
\label{eq:ffixs''_def2}
\end{equation}
Using the spherical identity equation (\ref{eq:I5}) and the same averaging approximations as before leads to
\begin{equation}
\vect[fix,s]{f}'' = \frac{3}{2} \phis \muf \lambdafix \Itens (\uf-\us)
\label{eq:ffixs''_def3}
\end{equation}
Substituting $\vect[fix,s]{f}'$ from equation (\ref{eq:ffixs'_def3}) and $\vect[fix,s]{f}''$ from equation (\ref{eq:ffixs''_def3}) into equation (\ref{eq:fjf_parts}) (with $j=\text{fix}$) yields after some simplification
\begin{equation}
\vect[fix,m]{f} = - \vnabla ( \phis \pf ) + \frac{5}{2} \muf \vnabla \cdot \left [ \phis \lambdafix \tensgammadotf \right ] - \vnabla \cdot \tens[slip,m]{\tau} + \vect[rot,m]{f}
\label{eq:ffixm_def}
\end{equation}
where
\begin{align}
\tens[slip,m]{\tau} & = - \frac{3}{2} \muf \phis (\vnabla \lambdafix) (\uf - \us) + \frac{3}{4} \muf \vnabla \left [ \phis \lambdafix (\uf-\us) \right ] \label{eq:tauslipm} \\
\vect[rot,m]{f} & = - 3 \muf \vnabla \cross \left [ \phis \lambdafix ( \omegaf - \omegas ) \right ] \label{eq:frotm}
\end{align}
This completes the derivation of $\vect[fix,m]{f}$.


\section{Interaction particle forces\label{sec:interaction_analysis}}

In this section we formulate expressions for the forces acting on the fluid and solid phases, $\vect[int,f]{f}$ ($=\vect[int,m]{f}-\vect[int,s]{f}$) and $\vect[int,s]{f}$, respectively, due to particle interactions.  These require evaluation of the interaction traction $\vect[int]{t}$ due to relative particle movement.

\subsection{Interaction particle stress model closure assumptions\label{sec:interaction_assumptions}}

Closure assumptions are required to model the interaction traction $\vect[int]{t}^p$ that acts on the surface of particle $p$.  Although this traction can be formerly defined from equation (\ref{eq:sigma_decomposition}), we model it using a pairwise sum over all surrounding particles ($q$) as
\begin{align}
\vect[int]{t}^p(\y) & = \vect[tot]{t}^p(\y,\uf,\us,\phis,\pf) - \vect[fix]{t}^p(\y,\uf,\us(\xp),\pf,\phis) \nonumber \\
& \approx \sum_{q \ne p} \vect[bin]{t}\left ( \np(\y),\tensgammadotbin(\xqp),\Deltaxqp \right )
\label{eq:tint_def}
\end{align}
where the mid-point between the pair of particles is $\xqp = (\xp+\xq)/2$ and $\Deltaxqp=\xq-\xp$ is the displacement between them.  In the above the pairwise interaction traction between two general particles has been defined as
\begin{equation}
\vect[bin]{t}(\n,\tensgammadot,\vect{r}) = \vect[bin*]{t}(\n,\tensgammadot,\vect{r}) - \vect[sin]{t}(\n,\tensgammadot)
\label{eq:tbin_def}
\end{equation}
where $\vect[bin*]{t}(\n,\tensgammadot,\vect{r})$ is the (total binary) traction acting on the surface at $\x + a \n$ of a particle at any general location $\x$ due to a binary interaction with a second particle existing at $\x + \vect{r}$ and where the two particles travel in an undisturbed strain rate field of $\tensgammadot$, and $\vect[sin]{t}(\n,\tensgammadot)$ is the equivalent (single) traction acting on the particle at $\x$ and surface at $\x + a \n$ and within the same undisturbed strain field of $\tensgammadot$, but when in isolation.  How $\tensgammadot$ is defined is detailed below.

A number of closure assumptions are employed in the approximation of equation (\ref{eq:tint_def}).  Firstly, the total traction difference represented by $\vect[tot]{t}^p-\vect[fix]{t}^p$ is approximated by a pairwise sum of traction differences between particle $p$ and each of its neighbours ($q$).  The `pairwise' contribution of $\vect[tot]{t}^p$ is modelled as $\vect[bin*]{t}$ in each term in this sum, representing the interaction between two particles in an infinite medium.  Similarly, the contribution to $\vect[fix]{t}^p$ in each term in the sum is modelled as $\vect[sin]{t}$, being the traction on an isolated particle in the same flow field and in an infinite medium which can be interpreted as the dilute limit of $\vect[fix]{t}^p$.  Note that as the distance between particles becomes large both $\vect[tot]{t}^p-\vect[fix]{t}^p$ and $\vect[bin]{t}$ approach a consistent zero limit.

The second closure assumption employed in equation (\ref{eq:tint_def}) is to ignore the dependence of $\vect[tot]{t}^p$ and $\vect[fix]{t}^p$ on the real fluid field variables $\uf$ and $\pf$, but instead define an undisturbed and divergence-free `interaction' velocity field in the neighbourhood of each particle interaction that is based on $\us(\xqp)$ and $\phis(\xqp)$.  Neglect of $\uf$ and $\pf$ on $\vect[int]{t}^p$ is justified given that the influence of these fluid fields is already captured in the total traction via $\vect[fix]{t}^p$, and that due to the linearity of the Stokes equations, the contribution to the total traction experienced by the reference particle due to these fields would only weakly depend on whether the surrounding particles were moving or not.  The interaction velocity field is defined by the strain rate tensor $\tensgammadotbin$, which is calculated for each binary interaction to ensure that the relative isolated velocity between the two particles is equal to $\us(\xq)-\us(\xp)$, but is also multiplied by a factor $\lambdaint$ that is a function of $\phis$ to account for multi-particle and even friction effects (in a phenomenological sense) acting during the interaction.  The calculated interaction velocity field is constrained to be divergence-free so that existing correlations for the force and force moment acting on the reference particle within an incompressible fluid during the interaction can be used.  More details of how $\tensgammadotbin$ is calculated from $\us$ and $\phis$ are given in section \ref{sec:interaction_bin}.  Note that as $\tensgammadotbin$ is evaluated using gradients of $\us$ and $\phis$ evaluated at $\xqp$, the same strain rate is used to evaluate the interaction traction between any pair of particles.

The next closure assumption involves replacing the sum over surrounding particles in the interaction traction equation (\ref{eq:tint_def}) with an integral over possible locations of the secondary particles.  To do this we first define integral moments of the tractions over the surface $\scali[\n]{S}$ of the general particle at $\x$ as
\begin{equation}
\vecti[j]{F}^{'^m}(\tensgammadot,\vect{r}) = \intSn {\n}^m \vecti[j]{t}(\n,\tensgammadot,\vect{r}) \dSn
\label{eq:Fbin_def}
\end{equation}
where the two arguments to $\vecti[j]{F}^{'^m}$ represent the characteristic interaction strain rate and relative interacting particle locations, respectively, and this equation applies for $j=$ bin, bin* or int and $m \ge 0$.  Note that due to equation (\ref{eq:tbin_def})
\begin{equation}
\vect[bin]{F}^{'^m}(\tensgammadot,\vect{r})=\vect[bin*]{F}^{'^m}(\tensgammadot,\vect{r})-\vect[sin]{F}^{'^m}(\tensgammadot)
\label{eq:Fbin_sum}
\end{equation}
Using equations (\ref{eq:tint_def}) and (\ref{eq:Fbin_def}) in equation (\ref{eq:fjs'm_def}) (with $j=$ int) leads to 
\begin{equation}
\vect[int,s]{f}^{'^m} = a^m \sum_p g(|\x-\xp|) \sum_{q \ne p} \vect[bin]{F}^{'^m}(\tensgammadotbin((\xqp),\Deltaxqp)
\label{eq:fsint_def3}
\end{equation}
We employ the particle density function (PDF) approach of (e.g.) \citet{batchelor72d} to replace the sum over surrounding particle ensembles in equation (\ref{eq:fsint_def3}) with an integral containing the probability $\mathcal{P}$ over the region surrounding the reference particle.  The dimensional PDF $\mathcal{P}$ can be expressed using a non-dimensional PDF $p$ via $\mathcal{P}=pn$ where $n$ is the particle averaged number density surrounding the reference particle, and the non-dimensional PDF $p \rightarrow 1$ as $|\vect{r}| = r$ becomes large \citep{batchelor72d}.  For our closure model we generalise \citeauthor{batchelor72d}'s approach further and allow the local PDF to depend on not only the distance between particle centres $r$, but also on the local divergence-free strain rate $\tensgammadots$, with this strain rate evaluated at the mid-point between the two particles.  Using this location ensures that the probability of an interaction occurring between two particular particles is equal for both.  Further details of how $p$ is modelled are given in section \ref{sec:interaction_pdf}.  With these assumptions the interaction forces become
\begin{align}
\vect[int,s]{f}^{'^m} = a^m \sum_p g(|\x-\xp|) &\intr n(\xp+\vect{r}) p(\tensgammadots(\xp+\vect{r}/2),\vect{r}) \times \nonumber \\
& \vect[bin]{F}^{'^m}(\tensgammadotbin(\xp+\vect{r}/2),\vect{r}) \dr
\label{eq:fsint_def4}
\end{align}
where like $\vect[bin]{F}^{'^m}$, the two arguments to $p$ represent the strain rate describing the particle interaction and relative particle locations, respectively.  Note that the integral over $d\vect{r}$ in equation (\ref{eq:fsint_def4}) is formerly taken over all space which could contain the centre of an interacting particle (i.e., $r>2a$), however, preempting the discussion of section \ref{sec:interaction_pdf}, in practice $p=0$ for $r>r_{\infty}$ (where $r_{\infty}<L$) so that this integral need only be performed between concentric spheres of radius $2a$ and $r_{\infty}$.

In the following sections \ref{sec:interaction_pdf}, \ref{sec:interaction_bin} and \ref{sec:interaction_Fm} we detail how $p$, $\tensgammadotbin$ and $\vect[bin]{F}^{'^m}$ are calculated, respectively, before deriving the final form of the interaction forces in section \ref{sec:interaction_forces}.

\subsection{Defining the interaction particle density function $p(\tensgammadot,\vect{r})$\label{sec:interaction_pdf}}

How the non-dimensional PDF $p$ is modelled is critical in determining the form of the suspension scale interaction stresses.  In particular, anisotropy of the PDF results in non-Newtonian suspension stresses, which cause shear induced migration of particles \citep{batchelor72d,lemaire23}.  Much research has been done on understanding the form of this PDF in solid-liquid suspension flows, with experimental measurements \citep{rampall97,gao10,blanc11b,blanc12a,blanc12} largely corroborating theoretical analysis \citep{arp77,zinchenko84,dacunha96,brady97,wilson00,wilson02,wilson05} and Stokesian simulations \citep{wilson02,morris02,drazer04,gao10,blanc12}.  Based on this knowledge, in the following we assume a general form for the PDF that is able to capture its main observable traits, while in paper II \citep{noori24a} we adopt a specific form for the PDF that is consistent with rough-particle interactions in strong flows, with coefficients chosen to reproduce experimentally observed particle migration at the suspension scale.

The general form adopted for $p$ is
\begin{equation}
p(\tensgammadots,\vect{r}) = \left [ \scali[2]{f}^*(r) - \scali[1]{f}^*(r) \tensgammahatdots : \rhat\rhat \right ] q (r)
\label{eq:pdf_def}
\end{equation}
where $q(r)$ is the isotropic PDF derived by \citet{batchelor72d} for dilute particle interactions, $\tensgammahatdots$ is the non-dimensional divergence-free strain rate that characterises the interaction (see equation (\ref{eq:gammadots_def})), $\rhat = \vect{r}/r$ is the unit binary interaction vector, $r=|\vect{r}|$ and $\scali[2]{f}^*(r)$ and $\scali[1]{f}^*(r)$ are functions of particle separation $r$ that describe the magnitudes of the isotropic and anisotropic regions within the PDF, respectively.  These $f^*$ functions are specified explicitly in paper II \citep{noori24a}, although we provide constraints on their values in the following.

The form of $p$ as represented by equation (\ref{eq:pdf_def}) has been chosen to capture four observed traits of the suspension PDFs:

\subsubsection{Isotropic PDF based on dilute particle interactions}

Considering only hydrodynamic pairwise interactions, \citeauthor{batchelor72d} found that for any point that lies on a streamline originating in the far field, $p(\vect{r}) = q(r)$, with $q(r)$ defined by
\begin{equation}
q(r) = \frac{1}{1-A}\exp{ \int^\infty_r \frac{3(B-A)}{r(1-A)} dr }
\label{eq:q_def}
\end{equation}
Here $A(r)$ and $B(r)$ are the same particle mobility functions used in Section \ref{sec:interaction_Fm}.  The proposed PDF uses $p=q$ in any isotropic regions where $\scali[2]{f}^*=1$ and $\scali[1]{f}^*=0$.

\subsubsection{Screening effects at large distances}

As highlighted by previous researchers \citep[e.g.][]{batchelor72d,drew76}, integrals of force moments such as contained in equation (\ref{eq:Fbin_def}) (with e.g. $m=1$) that are based on hydrodynamic pairwise interactions are mathematically divergent when evaluated over all space (i.e., all surrounding particles).  \citet{batchelor72d} used this as motivation to develop a renormalisation technique that uses an analogy between the local and average strain rates generated by the interacting particle to model the total force dipole on the reference sphere as that due to the closest interacting particle.  Later studies recognised that the renormalisation terms included in the dipole integrals do not (by design) affect their evaluation, provided that the integral is performed in polar coordinates, with the radial integration carried out last \citep{zinchenko84,wilson00}.

Our problem is more complex than these previous works in that we need integrals of both the pairwise force moments \emph{and} forces (multiplied by $\vect{r}$) evaluated over the surrounding particles, as represented by equation (\ref{eq:Fbin_def}) with $m=1$ and $m=0$, respectively.  Rather than trying to adapt \citet{batchelor72d}'s renormalisation technique, we build upon the observations of \citet{jackson97} and take a different approach.  Namely, we contend that there is no reason why the average forces or force moments exerted on a particle need be equal to the sum from pairwise interactions from all surrounding particles, given that the forces and moments from individual particles at larger distances from the reference sphere may be screened by those that are closer.  Hence, our pragmatic approach to evaluating the stress integral in equation (\ref{eq:Fbin_def}) is to use $p=0$ (i.e., $\scali[1]{f}^*=\scali[2]{f}^*=0$) for $r>\rinfty$.  We refer to $\rinfty$ as a `screening length'.

Note that other researchers have used a similar maximum PDF radius:  For example \citet{gillissen18} and \citet{gillissen19} who consider intermediate particle volume fractions and smooth particles, use an equivalent $r_\infty$ that is just larger than $2a$ to limit interactions to particles that are in close contact with the reference particle, lying in an `interaction shell'.

In similarity with the renormalisation technique, we do recognise that in a semi-dilute suspension forces and force moments on the reference particle will be dominated by those generated by the closest interacting particles, and hence set $\rinfty$ to be a function of the distance to the nearest neighbouring particle(s).  This produces integrated stresses that are convergent for $\phis \rightarrow 0$, despite $\rinfty \rightarrow \infty$ in the same dilute limit.  Details of how $\rinfty$ are evaluated are provided in paper II \citep{noori24a}.

\subsubsection{Exclusion of particles from the inner closed streamline region}

Under pure straining flow a region of closed streamlines surrounds the reference sphere, extending in both directions outwards from the centre of the reference sphere over parts of the velocity and vorticity plane \citep{batchelor72d,arp77}.  The probability of finding a second sphere within this closed streamline region (forming a doublet) is not represented by $q(r)$, but rather depends on an interplay between surface roughness, multi-body interactions and flow history.  A second sphere trapped within these closed streamlines, and rotating around the reference sphere close to the shear plane (ie, the plane formed by the velocity and velocity gradient vectors), must pass within a distance of $4.2\ee{-5}a$ or smaller from the reference sphere\citep{arp77}.  However, as this dimension is smaller than the roughness of most practical particles, contact forces push the second particle across the hydrodynamic streamlines, causing secondary spheres to be excluded from this part of the closed streamline region.  This mechanism has been confirmed via trajectory analysis \citep{dacunha96} and via examining PDFs measured from both experiments \citep{rampall97,blanc12,blanc11b} and Stokesian dynamics simulations\citep{drazer04,morris02,blanc12}.  In particular, the dilute ($\phi=5\%$) PDF measurements of \citet{blanc11b}, measured near the shear plane, show that the shape of the excluded particle region is consistent with theory, being fore-aft symmetric around the velocity gradient and vorticity plane and extending as a narrowing slit outwards from the reference particle in both directions along the velocity axis.

As the particle concentration increases, multi-body interactions become more important, and are able to move secondary particles both into and out of the closed streamline region.  The experimental PDF measurements of \citet{blanc12a} show that as the volume fraction increases from $\phi=5\%$, the regions of particle deficit (that is, $p(\vect{r}) < 1$) along the velocity axis and within the shear plane reduce in size, until at $\phi\approx25\%$, they are no longer visible at the resolution of the data.  This is consistent with the experimental results of \citet{rampall97} that show particle probability densities increasing within these excluded regions as the volume fraction is increased from $\phi=5\%$ to $15\%$, and with the Stokesian dynamics results of \citet{drazer04} that show that spherical averages of the PDF function across a range of radii approach the isotropic PDF $q(r)$ as the particle volume fraction increases from $\phi=1\%$ to $25\%$.  Similar results at higher volume fractions also show that the isotropic spherical average of $p(\vect{r})$ is well represented by $q(r)$, suggesting that the closed streamline particle deficit regions are absent at these volume fractions \citep{morris02}.

Note that closed streamlines surrounding the reference particle that are located away from the shear plane (that is, some distance along the vorticity axis away from the centre of the reference particle) have a larger minimum approach distance between spheres compared to those occurring on the shear plane\citep{arp77}.  This means that secondary particles impacting away from the shear plane can be caught in doublets and not excluded by practically relevant roughness.  Although there are few PDF measurements available in the literature which look beyond the shear plane, the Stokesian dynamics results of \citet{gao10} at $\phi=31.9\%$ and shown over the velocity and vorticity plane support this analysis.  Under dilute conditions it is hence difficult to predict the probability of finding particles within these outlying parts of the closed streamline region.  However for our purposes this is of minor significance because;  a) the results of \citet{arp77} show that the secondary particles must impact the reference particle quite close to the velocity and vorticity plane for the minimum separating gap size to be larger than practical particle roughness, and b) the effect of any inaccuracies from this part of the PDF on the resulting stress will be slight as the number of secondary particle impacts within this region under linear shear conditions is low, given the proximity to the same velocity and vorticity plane.

It should be noted that considerably less research has been conducted on the PDF formed during linear extensional strain, as opposed to linear shear strain.  \citet{batchelor72d} predict that no closed streamline regions are present during extensional strain, however it is expected that particle roughness will still cause secondary particle deficits in regions of extension, given that under extension strain colliding particles that are subject to hydrodynamic forces still pass within small distances of the reference sphere\citep{arp77}.

To capture the absence of particles occurring within the closed streamline inner region under our targeted semi-dilute conditions, we use an approximate model and assume that $\scali[1]{f}^*=\scali[2]{f}^*=0$ for $r<r_0$, where $r_0>2a$ is a critical inner region dimension defined in paper II \citep{noori24a}.

\subsubsection{Anisotropic PDF caused by particle roughness}

As well as excluding particles from the closed streamline region, particle roughness also alters the trajectories of secondary spheres that approach the reference sphere from the far field.  Under dilute conditions in a linear shear flow, this has the effect of moving the departing particle boundary of the thin excluded particle region further into the expansion quadrant of the flow\citep{dacunha96,rampall97}, breaking the fore-aft symmetry of the particle distribution (around the velocity gradient and vorticity plane)\citep{drazer04}.  As mentioned above, it is this fore-aft asymmetry that causes normal stresses to develop in the suspension.  As the particle volume fraction increases and the size of the excluded particle regions decreases, the particle distribution becomes more fore-aft asymmetric, with the location of maximum particle deficit near $r=2a$ moving around the reference sphere, towards the centre of the expansion quadrant\citep{blanc12}.  At high volume fractions ($\phi \gtrsim 45\%$) complex structures develop that are highly anisotropic, with clustering particularly at increasing radial multiples of the particle diameter\citep{gao10,blanc12}.  These structures presumably result from the interplay between particle roughness, multi-body interactions and possibly contact friction.

Following the above physical description of how roughness induces anisotropy in the PDF, we postulate that $p$ is also a function of the non-dimensional relative velocity between the two particles as represented by $\tensgammahatdots : \rhat\rhat$.  Indeed, in regions where this parameter is negative, secondary particles move from the far field towards the reference particle, and the local PDF may be increased or decreased (depending on $r$) due to roughness stopping the relative normal motion of the particles.  Conversely, in regions where $\tensgammahatdots : \rhat\rhat > 0$, particles move away from the reference particle, and realising that these particles may have been displaced from their approaching hydrodynamic streamlines by the particle roughness, the local PDF may be decreased relative to the isotropic $q(r)$ value.  Hence, on physical grounds, the parameter $\tensgammahatdots : \rhat\rhat$ is included in equation (\ref{eq:pdf_def}) as a means of modelling the expanding and contracting regions of the flow surrounding the reference particle.

The inclusion of $\tensgammahatdots : \rhat\rhat$ in the PDF form can also be justified based on previous experimental analysis.  Specifically, \citet{morris02} performed a spherical harmonic decomposition of the PDF obtained via Stokesian dynamics for the linear shear of a $\phi=55\%$ suspension, including in their simulations a small amount of Brownian diffusion ($\dnum{Pe}=1000$).  Of the first nine terms of the decomposition, the first ($g_{00}$ in their notation) dominates at all radial locations shown.  This term represents an isotropic contribution to the PDF and is captured by $q(r)$ in our equation (\ref{eq:pdf_def}).  The next two largest terms are represented by $g_{22}$ and $g_{2,-2}$.  The first of these ($g_{22}$) contributes an even function (around the velocity gradient and vorticity plane) to the PDF that hence does not contribute to the normal stresses \citep{batchelor72d}.  The second of these terms however ($g_{2,-2}$) contributes an odd function (around the velocity and vorticity plane) to the PDF, and thus is the largest term in the decomposition that contributes to the normal stresses in the suspension.  The harmonic basis function for this term can be shown to be equal to the anisotropic multiplier used in our PDF under linear shear conditions (that is, $\tensgammahatdots : \rhat\rhat$), implying that our $q(r)\scal[1]{f}^*(r)$ in equation (\ref{eq:pdf_def}) is conceptually equivalent to the $g_{2,-2}$ coefficient.  Hence, our proposed PDF can represent the isotropic and leading normal stress terms to this measured PDF when represented in spherical harmonics.

In principle we could represent the PDF using a greater number of spherical harmonic terms than used in equation (\ref{eq:pdf_def}), however to represent these in a frame invariant manner would require a significantly more mathematically complex PDF.  Also, as discussed by \citet{morris02}, it may take many terms for the series to converge. This is certainly an avenue for future work.

\subsection{Defining the binary interaction strain rate tensor, $\tensgammadotbin$ \label{sec:interaction_bin}} 

We next detail how the local binary interaction far-field fluid strain rate $\tensgammadotbin$ is calculated from $\us$ and $\phis$.  This tensor represents the strain rate of the incompressible fluid that locally surrounds each interacting pair of particles, and is chosen so that the particles, when moving force and torque free and in isolation, and at infinite dilution, move at the local velocity defined by $\us$.  As a sphere in linear Stokes flow travels at the velocity of the undisturbed fluid at its centroid location, this implies that the local fluid velocity of the reconstructed `dilute' velocity field is also equal to $\us$ at the centroid of both interacting particles, and that the velocity of the reference sphere relative to this reconstructed dilute velocity field is zero.

There are additional modelling constraints imposed on the calculation of $\tensgammadotbin$.  As the subsequent force analysis only requires the symmetrical and spherical parts of this strain-tensor, only these parts are included in $\tensgammadotbin$, with the anti-symmetric part of this flow field left undefined.  Further, in the absence of any deviatoric but presence of dilational strain in the $\us$ field, the dilute interaction velocity field is so chosen to be symmetrical around the axis between particles ($\rhat$), capturing the correct relative particle velocity as described by $\gammadotsphs$.  Note that unlike the particle scale velocity field used in the analysis of the fixed particle forces, for the interaction forces the representation of the suspension scale divergence at the local particle scale is not confined to the physical dimensions of the problem.  This modelling choice makes the following mathematical analysis tractable.  Finally, as detailed below, under non-dilute conditions the calculated `dilute' strain rate is increased to capture the effects of multi-particle interactions.  

With these constraints the following strain rate tensor results:
\begin{equation}
\tensgammadotbin = \lambdaint \left [ \tensgammadots + \frac{1}{2} \gammadotsphs \left ( 3 \rhat \rhat - \Itens \right ] \right ] 
\label{eq:tensgammadotbin_def}
\end{equation}
The first term in the square braces is the symmetrical divergence-free component of the solid velocity field, $\tensgammadots$, or deviatoric strain rate of the solid particles.  Note from equation (\ref{eq:tensgammadoti_def}) with $i=\text{s}$ that $\tensgammadots$ is already divergence-free.  The second term of $\tensgammadotbin$ captures the expansion and contraction (dilation) of the solid particle field, represented via $\gammadotsphs$ as defined via equation (\ref{eq:gammadotsphi_def}) with $i=\text{s}$.  And finally, like $\lambdafix$ for the fixed stresses, the `dilute' strain rate calculated by matching the local $\us$ field is multiplied by a factor $\lambdaint$ to increase the applied stresses occurring during an interaction, accounting for multi-particle interactions using an effective medium concept.  This factor can equivalently be viewed as an increase in viscosity acting during the interaction.  It is included in the calculation for the interaction strain rate $\tensgammadotbin$ so that it is evaluated at the mid-point between particles (like $\tensgammadotbin$), ensuring reciprocity of forces acting on each particle during each interaction.  

\subsection{Defining the binary interaction force $\vect[bin]{F}$ and force moment $\vect[bin]{F}'$ \label{sec:interaction_Fm}} 

With $\tensgammadotbin$ and $\vect[bin]{t}$ defined, we now outline a physical model that describes how the two particles hydrodynamically interact so that expressions for $\vect[bin]{F}$ and $\vect[bin]{F}'$ can be determined.

Focusing first on the force the fluid exerts on each particle during the interaction, $\vect[bin]{F}$ (defined by equation (\ref{eq:Fbin_def} with $m=0$), equation (\ref{eq:Fbin_sum}) gives $\vect[bin]{F}=\vect[bin*]{F}-\vect[sin]{F}$, where $\vect[bin*]{F}$ is the total force exerted during the pairwise interaction, and $\vect[sin]{F}$ is the force exerted when the particle travels in isolation.  Calculating $\vect[sin]{F}$ is straightforward, using known results for isolated spheres in a linear strain rate field.  Calculating $\vect[bin*]{F}$ is more complex however, requiring a model for the trajectory of the particles, or equivalently the net forces acting on them, during the interaction.  We discuss three models that describe the interaction --- the `constrained', `free' and `semi-free' models --- before choosing the last to implement mathematically to determine $\vect[bin*]{F}$.

The central assumption under the constrained model is that the particles move exactly at the local reconstructed fluid velocity (i.e., based on $\tensgammadotbin$) during an interaction, or equivalently, that the slip velocity of each particle is at all times zero.  \citet[p178]{kim91} provides a resistance matrix that expresses the forces, torques and stresslets that the fluid exerts on two interacting particles within a Stokesian linear strain field as a function of the each of the particle's slip velocities and slip rotations, as well as the undisturbed fluid deviatoric strain tensor.  As we shall see in more detail below, we are primarily interested in finding the forces and stresslets that the fluid exerts on the particles in order to calculate $\vect[bin]{F}$ and $\vect[bin]{F}'$, respectively.  Hence, in principal the constrained method involves setting the slip velocity of each particle to zero (and potentially the slip rotation too), and then using the binary strain rate tensor $\tensgammadotbin$ to calculate the force, torque and stresslet exerted by the fluid on the particles from the available resistance matrix.

The first complication with this approach is that not all particle trajectories are physically feasible under the constrained approach.  Specifically, if the surrounding interacting particle trajectory comes within a distance of $2a$ from the reference particle (i.e., $r< 2a$) the particles will overlap, and the resistance matrix coefficients become undefined.  As we shall find, particles that come within small distances of each other are particularly important to calculating the suspension stress tensors, so this deficiency of the constrained method is significant.  A second issue with the constrained method is that for these particle dynamics to be realised in a multi-particle system, the net force from the remaining particles would have to be sufficient to completely resist the force from the binary interaction.  This is a poor assumption, particularly for interacting particles that are closest to the reference particle and hence dominate the generated particle stresses.

By contrast, the central assumption under the free method is that during an interaction each particle is force and torque free.  This resolves the overlapping trajectory problem of the constrained method as under the free method particle trajectories do not exactly follow the undisturbed fluid streamlines, but rather deviate from these streamlines according to the stokes flow solution, moving around each other as they pass by (in the case of deviatoric strain).  However, due to the (net) force free assumption under the free model, each particle within a binary interaction feels no resistance to movement from the remaining particles.  Noting that each pairwise interaction actually occurs within a field of other particles, this is a poor assumption, particularly as the particle concentration increases.  Note that under the free model, $\vect[bin*]{F}=0$, leading (after subsequent analysis) to $\tauints=0$ and no interaction induced particle migration.

As a compromise, in this study we use a `semi-free' model of interaction that calculates the resistance offered by the remaining particles during a binary interaction as intermediate between the constrained and free models.  It also avoids the $r<2a$ issue of the constrained model.  Under this model the trajectories of the interaction particles are calculated using the free model, but then the force that the interacting particle exerts on the reference particle is modelled as that would be required on the reference particle, when in isolation, that causes it to move at the relative velocity predicted from the two particle problem.  A similar assumption is used for the torque that the fluid exerts on each particle during the interaction.  To find the trajectories of the interacting particles, we use the two particle mobility matrix provided by \citet[p179]{kim91} that expresses the particle slip velocities, particle slip rotations, and particle stresslets in terms of the net forces and torques acting on each particle, as well as the undisturbed deviatoric strain tensor.  So by setting the force and torque acting on each particle to zero, the slip velocity and slip rotation of both particles can be calculated.

To calculate the value of $\vect[sin]{F}$ under the semi-free model, we recognise that the local undisturbed velocity field defined by $\tensgammadotbin$ was constructed so that when in isolation the reference particle moves at the local solid velocity value, implying no slip velocity and no net force on the single isolated particle.  Hence $\vect[sin]{F}=\vect{0}$.  To calculate the value of $\vect[bin*]{F}$ under the semi-free model, we recognise that the effective interaction force is balanced by a drag force that also acts on the particle, and has value of $\vect[drag]{F} = - 6 \pi a \muf \vect{\Delta u}$ for a spherical particle moving in a Stokesian linear strain field \citep[e.g. see][]{kim91}.  Hence $\vect[bin*]{F} = - \vect[drag]{F} = 6 \pi a \muf \vect{\Delta u}$.  To calculate $\vect{\Delta u}$, which is the velocity of the reference particle during the binary collision relative to the undisturbed fluid velocity, we set the net force and torque acting on both particles to zero in the two particle mobility matrix \citep[i.e. p179 in][]{kim91}, giving $\vect{\Delta u} = - \frac{1}{2} \tensi[1]{\widetilde{g}} : \tensgammadotbin$, where $\tensi[1]{\widetilde{g}}$ is a mobility tensor defined in \citeauthor{kim91}.  Expanding this tensor using the relationships in \citet{kim91} for two equally sized spherical particles leads to 
\begin{equation}
\vect[bin]{F}(\tensgammadotbin,\vect{r}) = \frac{3}{2} \pi a \muf r \left [ A \rhat \rhat + B \left ( \Itens - \rhat \rhat \right ) \right ] \rhat : \tensgammadotbin
\label{eq:Fbin_def4}
\end{equation}
In this expression $A(r)$ and $B(r)$ are the same two mobility functions as used by \citet{batchelor72d} in the definition of the PDF $q$ in equation (\ref{eq:q_def}).  Specific definitions of these mobility functions are given by equations (\ref{eq:A_def}) and (\ref{eq:B_def}) in terms of secondary mobility functions that are defined for near and far field interactions by \citeauthor{kim91}.

Turning now to the force moment on the reference particle $\vect[bin*]{F}'$: the free interaction model interpretation of this quantity is of that which the reference particle experiences during a force free collision, while the semi-free interpretation of this could differ by at most the force moment associated with a particle moving at $\Delta \vect{u}$ in an infinite medium.  However, for an isolated spherical particle travelling in a linear strain rate field the stresslet is independent of the particle slip velocity \citep{kim91}, so this latter contribution is zero and the free and semi-free interaction models result in the same evaluation of $\vect[bin*]{F}'$ that has been used in many previous works \citep{batchelor72d,zinchenko84,wilson00} (albeit, without inclusion of the isotropic component).

To calculate this force moment in the present context, we relate the force moment as defined by equation (\ref{eq:Fbin_def}) with $m=1$ to \citet{kim91}'s force dipole $\tens{D}$ definition as
\begin{equation}
a ( \vect{F}' )^T = \tens{D} = \tens{S} + \tens{T} + \frac{1}{3} \scal[tr]{S} \Itens
\label{eq:force_dipole}
\end{equation}
Further, as indicated this force dipole can be expressed as the sum of a symmetric or deviatoric `stresslet' $\tens{S}$, an antisymmetric tensor $\tens{T}$ and an isotropic `pressure moment' component which we define as $\scal[tr]{S}=\trace{\tens{D}}$ \citep{jeffrey93}.  The antisymmetric tensor $\tens{T}$ is directly related to the torque exerted on the particle \citep[e.g.][]{kim91}, and as the reference sphere is defined to be torque free during both its interaction with the surrounding sphere (for $\vect[bin*]{F}'$) and when in isolation (for $\vect[sin]{F}'$) this tensor can be neglected.  Noting the symmetry of the remaining components, equations (\ref{eq:Fbin_sum}) and (\ref{eq:force_dipole}) can be combined to give
\begin{equation}
\vect[bin]{F}'(\tensgammadotbin,\vect{r}) = \frac{1}{a} \left [ \tens[bin*]{S} - \tens[sin]{S} + \frac{1}{3} \left ( \scal[tr,bin*]{S} - \scal[tr,sin]{S} \right ) \Itens \right ]
\label{eq:Fbin'_def3}
\end{equation}
with the subscripts bin* and sin referring to particle properties associated with the reference particle during the binary collision and in isolation, respectively, with both evaluated within the velocity field described by $\tensgammadotbin$.  

The deviatoric stresslets $\tens[bin*]{S}$ and $\tens[sin]{S}$ can be evaluated in a straight-forward manner using the mobility matricies given in \citet{kim91}.  Specifically, for spherical force and torque free particles we have
\begin{align}
\tens[bin*]{S} - \tens[sin]{S} = & \frac{\muf}{2} \left ( \tensi[1]{m} - \tens{M} \right ) : \tensgammadotbin \nonumber \\
= & \frac{10}{3}\pi \muf a^3 \biggl\{ K \tensgammadotbin + \left ( \tensgammadotbin \cdot \rhat \rhat + \rhat \tensgammadotbin \cdot \rhat \right ) L + \nonumber \\
& \tensgammadotbin : \rhat \rhat \left [ \rhat \rhat M - \left ( \frac{2}{3}L + \frac{1}{3} M \right ) \Itens \right ] \biggr\}
\label{eq:Sbin*_def}
\end{align}
where $\tensi[1]{m}$ and $\tens{M}$ are mobility tensors defined for binary interactions and particles in isolation, respectively, and where the mobility functions $K(r)$, $L(r)$ and $M(r)$ are defined using equations (\ref{eq:K_def}), (\ref{eq:L_def}) and (\ref{eq:M_def}), respectively, in terms of secondary mobility functions tabulated for near and far field interactions in \citet{kim91}.

The isotropic or pressure moment component of the force dipole is provided by \citet{jeffrey93}.  The isolated sphere moment $\scal[tr,sin]{S}$ is proportional to the undisturbed pressure at the sphere location \citep{jeffrey93}, however noting that the undisturbed velocity field is described by a linear strain rate $\tensgammadotbin$, and that the effects of the pressure field on the particle stress are already captured in $\tfixp$, we use a uniform and zero undisturbed pressure field during the interaction modelling giving $\scal[tr,sin]{S} = 0$.  To calculate $\scal[tr,bin*]{S}$ \citet{jeffrey93} provides a two particle resistance tensor that is a function of the slip velocities of both particles, and the undisturbed fluid strain rate.  Noting from previous analysis that the slip velocity of the reference particle relative to the undisturbed velocity at its location is $\vect{\Delta u} = - \frac{1}{2} \tensi[1]{\widetilde{g}} : \tensgammadotbin$, that the slip velocity of the interacting particle is the negative of this (as $\tensi[1]{\widetilde{g}} : \tensgammadotbin = -\tensi[2]{\widetilde{g}} : \tensgammadotbin$ using the notation of \citeauthor{kim91}), and that the deviatoric strain rate within each solid particle is zero, the pressure moment during the interaction can be expressed as
\begin{align}
\scal[tr,bin*]{S} & = \frac{\muf}{2} \left [ \left ( \vecti[11]{P} - \vecti[12]{P} \right ) \cdot \tensi[1]{\widetilde{g}} + \tensi[11]{Q} + \tensi[12]{Q} \right ] : \tensgammadotbin \nonumber \\
& = 10\pi \muf a^3 \left ( Q - PA \right ) \rhat \rhat : \tensgammadotbin
\label{eq:Strbin*_def}
\end{align}
where $\vecti[11]{P}$, $\vecti[12]{P}$, $\tensi[11]{Q}$ and $\tensi[12]{Q}$ are resistance vectors and tensors defined in \citet{jeffrey93}, and $P(r)$, $Q(r)$ and the previously used $A(r)$ are mobility and resistance functions defined via equations (\ref{eq:P_def}), (\ref{eq:Q_def}) and (\ref{eq:A_def}), respectively.  \citet{jeffrey93} provides near and far field values for $P$ and $Q$.

Finally, combining equations (\ref{eq:Fbin'_def3}) (\ref{eq:Sbin*_def}) and (\ref{eq:Strbin*_def}) gives the force moment during each pairwise interaction as
\begin{multline}
\vect[bin]{F}'(\tensgammadotbin,\vect{r}) = \frac{10}{3}\pi \muf a^2 \Bigl[ \tensgammadotbin K + \left ( \tensgammadotbin \cdot \rhat \rhat + \rhat \tensgammadotbin \cdot \rhat \right ) L + \\
\tensgammadotbin : \rhat \rhat \left ( \rhat \rhat M + N \Itens \right ) \Bigl]
\label{eq:Fbin'_def4}
\end{multline}
where $N(r)$ is a compound mobility function defined by equation (\ref{eq:N_def}).  This expression is equivalent to that used in previous works \citep{batchelor72d,zinchenko84,wilson00}, except for the isotropic terms that are contained in $N$ and that end up making contributions to the normal and dilational mixture viscosities of the suspension.

\subsection{Form of the particle interaction forces\label{sec:interaction_forces}}

With $p$, $\tensgammadotbin$ and $\vect[bin]{F}^{'^m}$ defined, the next objective is to remove the particle averaging from equation (\ref{eq:fsint_def4}) and in so doing express the functions $n$, $p$ and $\vect[bin]{F}^{'^m}$ in terms of suspension averaged properties evaluated at $\x$ (the location of $\vect[int,s]{f}^{'^m}$), rather than at $\xp+\vect{r}/2$ or $\xp+\vect{r}$.

To simplify subsequent manipulations we first define
\begin{equation}
\tens{h}^{'^m}(\xp+\vect{r}/2,\vect{r}) = p(\tensgammadots(\xp+\vect{r}/2),\vect{r}) \vect[bin]{F}^{'^m}(\tensgammadotbin(\xp+\vect{r}/2),\vect{r})
\label{eq:h_def}
\end{equation}
where the first argument of $\tens{h}^{'^m}$ specifies where the respective interaction strain rates should be evaluated, and the second the relative location of particles.  Using this definition equation (\ref{eq:fsint_def4}) becomes
\begin{equation}
\vect[int,s]{f}^{'^m} = a^m \sum_p g(|\x-\xp|) \intr n(\xp+\vect{r}) \tens{h}^{'^m}(\xp+\vect{r}/2,\vect{r}) \dr
\label{eq:fsint_def5}
\end{equation}
Expanding the location in which suspension properties are evaluated in $\tens{h}^{'^m}$ around $\xp$ using a Taylor series gives
\begin{equation}
\tens{h}^{'^m}(\xp+\vect{r}/2,\vect{r}) = \tens{h}^{'^m}(\xp,\vect{r}) + \frac{\vect{r}}{2} \cdot \vnablaxp \tens{h}^{'^m}(\xp,\vect{r}) + \order{\frac{\rinfty^2}{L^2}\vect[bin]{F}^{'^m}}
\label{eq:htaylor}
\end{equation}
given that $\order{p}=1$ only for $r<r_\infty$, and that suspension properties (that determine $\tens{h}^{'^m}$) vary over the lengthscale $L$.  Similarly, expanding $n(\xp+\vect{r})$ around $\xp$ using a Taylor series gives
\begin{equation}
n(\xp+\vect{r}) = n(\xp) + \vect{r} \cdot \vnablaxp n(\xp) + \order{\frac{\rinfty^2}{L^2}n}
\label{eq:ntaylor}
\end{equation}
Combining these gives after some manipulation
\begin{align}
n(\xp+\vect{r}) & \tens{h}^{'^m}(\xp+\vect{r}/2,\vect{r}) = n(\xp) \tens{h}^{'^m}(\xp,\vect{r}) + n(\xp) \frac{\vect{r}}{2} \cdot \vnablaxp \tens{h}^{'^m}(\xp,\vect{r}) \nonumber \\
& \quad + \left [ \vect{r} \cdot \vnablaxp n(\xp) \right ] \tens{h}^{'^m}(\xp,\vect{r}) + \order{\frac{\rinfty^2}{L^2}n\vect[bin]{F}^{'^m}} \nonumber \\
& = n(\xp) \tens{h}^{'^m}(\xp,\vect{r}) + \nonumber \\
& \quad + \vect{r} \cdot \frac{1}{2n(\xp)} \vnablaxp \left [ n^2(\xp) \tens{h}(\xp,\vect{r}) \right ] + \order{\frac{\rinfty^2}{L^2}n\vect[bin]{F}^{'^m}}
\label{eq:nhtaylor}
\end{align}
Using this expression in equation (\ref{eq:fsint_def5}), noting that any properties that are evaluated at $\xp$ can be removed from the integral over $\vect{r}$, gives
\begin{align}
\vect[int,s]{f}^{'^m} & = a^m \sum_p g(|\x-\xp|) n(\xp) \intr \tens{h}^{'^m}(\xp,\vect{r}) \dr \nonumber \\
 & + a^m \sum_p g(|\x-\xp|) \frac{1}{2n(\xp)} \vnablaxp \cdot \left [ n^2(\xp) \intr \vect{r} \tens{h}^{'^m}(\xp,\vect{r}) \dr \right ] \nonumber \\
& + \order{a^m \frac{\rinfty^5}{L^2}n^2\vect[bin]{F}^{'^m}}
\label{eq:fsint_def6}
\end{align}
Applying the particle phase averaging to all properties evaluated at $\xp$, and using the repeated averaging approximations from Appendix \ref{sec:averaging_repeated} gives
\begin{align}
\vect[int,s]{f}^{'^m} & = a^m n^2(\x) \intr \tens{h}^{'^m}(\x,\vect{r}) \dr \nonumber \\
 & + a^m \frac{1}{2} \vnablax \cdot \left [ n^2(\x) \intr \vect{r} \tens{h}^{'^m}(\x,\vect{r}) \dr \right ] + \order{a^m \frac{\rinfty^5}{L^2}n^2\vect[bin]{F}^{'^m}}
\label{eq:fsint_def7}
\end{align}
Reapplying the definition of $\tens{h}^{'^m}$ from equation (\ref{eq:h_def}), and dropping specific reference to $\x$ gives
\begin{align}
\vect[int,s]{f}^{'^m} & = a^m n^2 \intr p(\tensgammadots,\vect{r}) \vect[bin]{F}^{'^m}(\tensgammadotbin,\vect{r}) \dr \nonumber \\
& + a^m \frac{1}{2} \vnabla \cdot \left [ n^2 \intr \vect{r} p(\tensgammadots,\vect{r}) \vect[bin]{F}^{'^m}(\tensgammadotbin,\vect{r}) \dr \right ] 
+ \order{a^m \frac{\rinfty^5}{L^2}n^2\vect[bin]{F}^{'^m}}
\label{eq:fsint_def8}
\end{align}
where it is understood that the strain rates $\tensgammadots$ and $\tensgammadotbin$ are now evaluated at $\x$.

\subsection{Defining the particle interaction stresses\label{sec:interaction_stresses}}

As described by equation (\ref{eq:fjf_parts}) with $j=$int, calculating $\vect[int,s]{f}$ and $\vect[int,m]{f}$ involves calculating $\vect[int,s]{f}^{'^m}$ for $m=0,1\text{ and }2$.  

Starting with $\vect[int,s]{f}$ (that is, equation (\ref{eq:fsint_def8}) with $m=0$), $p$, as described by equation (\ref{eq:pdf_def}), contains only terms that have even multiples of $\rhat$, while $\vect[bin]{F}$, as described by equation (\ref{eq:Fbin_def}), contains only terms that have odd multiples of $\rhat$.  As spherical integrals involving odd multiples of normals evaluate to zero (see Appendix \ref{sec:integrals}), the first term on the right of equation (\ref{eq:fsint_def8}) makes no contribution to $\vect[int,s]{f}$, and hence the entire force can be written as the divergence of a stress,
\begin{equation}
\vect[int,s]{f} = - \vnabla \cdot \tens[int,s]{\tau} + \order{\frac{\rinfty^5}{L^2}n^2\vect[bin]{F}}
\label{eq:fints_def}
\end{equation}
where
\begin{equation}
\tens[int,s]{\tau} = -  \frac{\phis^2}{2 \nu^2} \intr \vect{r} p(\tensgammadots,\vect{r}) \vect[bin]{F}(\tensgammadotbin,\vect{r}) \dr
\label{eq:tauints_def}
\end{equation}
and the relative error of $\order{a^2/L^2}$ associated with equation (\ref{eq:aveidphis}) has been neglected.  The error in equation (\ref{eq:fints_def}) is $\order{\rinfty^2/L^2}$ smaller than $\vect[int,s]{f}$ and $\order{\rinfty^2/(aL)}$ smaller than $\vect[int,m]{f}$, so can be neglected provided that the domain size satisfies $L \gg \rinfty^2/a$.  Substituting the above equations for $p$ and $\vect[bin]{F}$ into equation (\ref{eq:tauints_def}) then gives
\begin{align}
\tens[int,s]{\tau} & = - \frac{27 \muf \phis^2 \lambdaint}{64 \pi a^5} \intr r^2 q \left ( \scali[2]{f}^* - \tensgammahatdots : \rhat\rhat \scali[1]{f}^* \right ) \times \nonumber \\
& \quad \rhat \left [ ( A - B ) \rhat \rhat + B \Itens \right ] \rhat : \left [ \tensgammadots + \frac{1}{2} \gammadotsphs \left ( 3 \rhat \rhat - \Itens \right ) \right ] \dr \nonumber \\
& = - \frac{27 \muf \phis^2 \lambdaint}{64 \pi a^5} \biggl \{
\intr r^2 q \scali[2]{f}^* (A-B) \tensi[1]{I} dr
+ \intr r^2 q \scali[2]{f}^* B \tensi[2]{I} dr
\nonumber \\ & \quad
- \intr r^2 q \scali[1]{f}^* (A-B)  \tensi[3]{I} dr
- \intr r^2 q \scali[1]{f}^* B  \tensi[4]{I} dr
\nonumber \\ & \quad
+ \gammadotsphs \Bigl [
\frac{3}{2} \intr r^2 q \scali[2]{f}^* A \tensi[5]{I} dr
- \frac{3}{2\gammadots} \intr r^2 q \scali[1]{f}^* A \tensi[1]{I} dr
\nonumber \\ & \quad
- \frac{1}{2} \intr r^2 q \scali[2]{f}^* A \tensi[5]{I} dr
+ \frac{1}{2\gammadots} \intr r^2 q \scali[1]{f}^* A \tensi[1]{I} dr \Bigr ] \biggr \}
\label{eq:tauints_def2}
\end{align}
where the spherical identities such as $\tensi[1]{I}$, $\tensi[2]{I}$, etc are defined and evaluated in section \ref{sec:integrals}, and where relevant in equation (\ref{eq:tauints_def2}), are evaluated using $\tensgammadots$.  Substituting the evaluations of these identities leads to
\begin{multline}
\tens[int,s]{\tau} = - \frac{27}{16} \muf \phis^2 \lambdaint \biggl \{ \Bigl [ \frac{2}{15} \left ( \chat{A2}-\chat{B2} \right ) + \frac{1}{3} \chat{B2} \Bigr ] \tensgammadots \\
- \gammadots \Bigl [ \left ( \frac{8}{105} ( \chat{A1} - \chat{B1} ) + \frac{2}{15} \chat{B2} \right ) \tensgammahatdots \cdot \tensgammahatdots + \frac{4}{105} ( \chat{A1} - \chat{B1} ) \Itens \Bigr ] \\
+ \gammadotsphs \Bigl [ \frac{1}{3} \chat{A2} \Itens - \frac{2}{15} \chat{A1} \tensgammahatdots \Bigr ] \biggr \}
\label{eq:tauints_def3}
\end{multline}
where the $\chat{Xi}$ are defined by equation (\ref{eq:chat_def}) and are integrals of the mobility coefficients ($A$ and $B$) and PDF $\scali[i]{f}^*$ functions.  Combining like terms leads to the general interaction stress equation (\ref{eq:tau_def4}) as applied to the solid phase ($i=\text{s}$), that contains the five non-dimensionalised viscosity functions, $\muints$, $\musidones$, $\musidtwos$, $\mubulkones$ and $\mubulktwos$, defined via equations (\ref{eq:muints}) to (\ref{eq:mubulktwos}), respectively.

Turning to $\vect[int,m]{f}$, applying equation (\ref{eq:fjf_parts}) with $j=\rm{int}$ gives
\begin{equation}
\vect[int,m]{f} = \vnabla \cdot \tens[int,s]{f}' - \frac{1}{2} \vnabla \cdot \left ( \vnabla \cdot \tens[int,s]{f}'' \right ) + \order{\frac{a^3 \rinfty^4}{L^4} n^2\vect[bin]{F}}
\label{eq:fintm_def}
\end{equation}
where the magnitude of $\vect[int,s]{f}$ has been evaluated based on equation (\ref{eq:fints_def}).  Similar to $\vect[int,s]{f}$, $\vect[int,s]{f}'$ and $\vect[int,s]{f}''$ are defined using equation (\ref{eq:fsint_def8}) evaluated with $m=1$ and $m=2$, respectively.  Noting from equation (\ref{eq:Fbin'_def4}) that $\vect[bin]{F}'$ only contains terms having even multiples $\rhat$, the second line of equation (\ref{eq:fsint_def8}) makes no contributions to $\vect[int,s]{f}'$, and we find
\begin{equation}
\vect[int,s]{f}' = a n^2 \intr p(\tensgammadots,\vect{r}) \vect[bin]{F}'(\tensgammadotbin,\vect{r}) \dr + \order{a \frac{\rinfty^5}{L^2}n^2\vect[bin]{F}}
\label{eq:f'sint_def}
\end{equation}
and
\begin{equation}
\vect[int,s]{f}'' = \order{a^2 \rinfty^3 n^2\vect[bin]{F}}
\label{eq:f''sint_def}
\end{equation}
The order of magnitude expressed by equation (\ref{eq:f''sint_def}) has originated from the first term in equation (\ref{eq:fsint_def8}) (evaluated with $m=2$), and in both equations $\order{\vect[bin]{F}^{'^m}}=\order{\vect[bin]{F}}$ has been applied based on equation (\ref{eq:Fbin_def}).  Noting that to leading order $\vect[int,s]{f}' = \order{a\rinfty^3n^2\vect[bin]{F}}$, substituting $\vect[int,s]{f}'$ and $\vect[int,s]{f}''$ back into equation (\ref{eq:fintm_def}) finally results in
\begin{equation}
\vect[int,m]{f} = - \vnabla \cdot \tens[int,m]{\tau} + \order{a \frac{\rinfty^5}{L^3}n^2\vect[bin]{F}}
+ \order{a^2 \frac{\rinfty^3}{L^2}n^2\vect[bin]{F}} 
\label{eq:fintm_def2}
\end{equation}
where
\begin{equation}
\tens[int,m]{\tau} = -  \frac{\phis^2 a}{\nu^2} \intr p(\tensgammadots,\vect{r}) \vect[bin]{F}'(\tensgammadotbin,\vect{r}) \dr
\label{eq:tauintm_def}
\end{equation}
and the two error terms in equation (\ref{eq:fintm_def2}) have originated from the truncation error associated with $\vect[int,s]{f}'$ as given in equation (\ref{eq:f'sint_def}), and that associated with $\vect[int,s]{f}''$ as given in equation (\ref{eq:f''sint_def}), respectively.  These errors are $\order{\rinfty^2/L^2}$ and  $\order{a/L}$ smaller than $\vect[int,m]{f}$, respectively, and are neglected.  

To find the final expression for the mixture interaction stress, we substitute the aforementioned equations for $p$ and $\vect[bin]{F}'$ into equation (\ref{eq:tauintm_def}), giving
\begin{multline}
\tens[int,m]{\tau} = - \frac{15 \muf \phis^2 \lambdaint}{8 \pi a^3} \intr r^2 q \left ( \scali[2]{f}^* - \tensgammahatdots : \rhat\rhat \scali[1]{f}^* \right ) \\
\Bigl\{ K \tensgammadots + 2L \tensgammadots \cdot \rhat \rhat + \left ( M \rhat \rhat + N \Itens \right ) \tensgammadots : \rhat \rhat + \\
\frac{\gammadotsphs}{2} \left [ ( 3K+4L+2M ) \rhat \rhat + (2N-K) \Itens \right ] \Bigr\} \dr
\label{eq:tauintm_def2}
\end{multline}
Expanding the multiplication, and then applying the spherical identities of section \ref{sec:integrals} results in
\begin{multline}
\tens[int,m]{\tau} = - \frac{15}{2} \muf \phis^2 \lambdaint \biggl \{ \left ( \dhat{K2}+\frac{2}{3}\dhat{L2}+\frac{2}{15}\dhat{M2} \right ) \tensgammadots \\
- \gammadots \Bigl [ \left ( \frac{4}{15} \dhat{L1} + \frac{8}{105} \dhat{M1} \right ) \tensgammahatdots \cdot \tensgammahatdots + \left ( \frac{4}{105} \dhat{M1} + \frac{4}{15} \dhat{N1} \right ) \Itens \Bigr ] \\
+ \gammadotsphs \Bigl [ \left ( \frac{1}{6} \dhat{R2} + \dhat{N2} - \frac{1}{2} \dhat{K2} \right ) \Itens - \frac{1}{15} \dhat{R1} \tensgammahatdots \Bigr ] \biggr \}
\label{eq:tauintm_def3}
\end{multline}
where similar to $\chat{Xi}$, $\dhat{Xi}$ are defined by equation (\ref{eq:dhat_def}) and are integrals of the mobility coefficients ($L$, $K$, $M$ and $S$) and PDF $\scali[i]{f}^*$ functions.  Collecting like terms again leads to the general interaction stress equation (\ref{eq:tau_def4}), but now applied to the mixture phase ($i=\text{m}$), that again contains five non-dimensionalised viscosity functions, being $\muintm$, $\musidonem$, $\musidtwom$, $\mubulkonem$ and $\mubulktwom$, defined via equations (\ref{eq:muintm}) to (\ref{eq:mubulktwom}), respectively.


\begin{figure}
\centering
\fbox{
\begin{minipage}{1.0\textwidth}
\center
\emph{Momentum equations:}
\begin{multline}
\rhof \left [ \frac{\partial \phif \uf }{\partial t}  + \vnabla \cdot \phif \uf \uf \right ] = - \phif \vnabla \pf - \vect[drag,s]{f} - \vect[faxen,s]{f} \\
- \vnabla \cdot \left ( \taudilm + \tauintm - \tauints \right ) + \rhof \phif \g
\label{eq:momentumf2}
\end{multline}%
\begin{multline}
\rhos \left [ \frac{\partial \phis \us }{\partial t}  + \vnabla \cdot \phis \us \us \right ] = - \phis \vnabla \pf + \vect[drag,s]{f} + \vect[faxen,s]{f}
- \vnabla \cdot \tauints + \rhos \phis \g
\label{eq:momentums2}
\end{multline}%
\begin{equation}
\frac{\partial \rhom \uvectrho }{\partial t}  + \vnabla \cdot \rhom \uvectrho \uvectrho = - \vnabla \pf - \vnabla \cdot \left ( \taudilm + \tauintm \right ) + \rhom \g
\label{eq:momentumm2}
\end{equation}

\vspace{1em}
\emph{Interphase force definitions:}
\begin{gather}
\vect[drag,s]{f} = \frac{9}{2a^2} \phis\muf\lambdafix(\uf-\us) \label{eq:fdrags} \\
\vect[faxen,s]{f} = \phis \muf \left \{ \frac{7}{4} \lambdafix \vnabla \cdot \left [ \tensgammadotf - (N-1) \ItensD \gammadotsphf \right ] + \frac{5}{2} \vnabla \lambdafix \cdot \tensgammadotf \right \} \label{eq:ffaxens}
\end{gather}

\vspace{1em}
\emph{Stress definitions:}
\begin{equation}
\taudilm = - \muf \left ( \tensgammadotm + \frac{5}{2} \phis \lambdafix \tensgammadotf \right )
\label{eq:taudilm_def}
\end{equation}%
\begin{align}
\tauinti = & - \muf \phis^2 \lambdaint \left [ \muinti \tensgammadots - \gammadots \left ( \musidonei \tensgammahatdots \cdot \tensgammahatdots + \musidtwoi \Itens \right ) \right . \nonumber\\
& \qquad \left . + \gammadotsphs \left ( \mubulkonei \Itens - \mubulktwoi \tensgammahatdots \right ) \right ] \qquad \text{for $i$ = s or m}
\label{eq:tau_def4}
\end{align}

\vspace{1em}
\emph{Strain rate definitions:}
\begin{equation}
\tensgammadotm = \vnabla \um + (\vnabla\um)^\text{T}
\label{eq:tensgammadotm_def}
\end{equation}
\begin{equation}
\gammadotsphi = \frac{2}{N} \vnabla \cdot \ui
\qquad \text{for $i$ = f or s}
\label{eq:gammadotsphi_def}
\end{equation}
\begin{equation}
\tensgammadoti = \vnabla \ui + ( \vnabla \ui )^T - \gammadotsphf \ItensD
\qquad \text{for $i$ = f or s}
\label{eq:tensgammadoti_def}
\end{equation}
\begin{equation}
\gammadots = \sqrt{\frac{1}{2} \tensgammadots : \tensgammadots} \qquad \tensgammahatdots = \frac{\tensgammadots}{\gammadots}
\label{eq:gammadots_def}
\end{equation}
\end{minipage}
}
\caption{Summary of simplified momentum equations, interphase forces, stresses and strain rate definitions.  As detailed in Appendix \ref{sec:dimensional_operator}, $N$ is the number of physical dimensions and $\ItensD$ is the unit tensor in those physical dimensions.\label{fig:summary_equations}}
\end{figure}


\begin{figure}
\fbox{
\begin{minipage}{1.0\textwidth}
\center
\emph{Interaction viscosities:}
\begin{gather}
\muints = \frac{9}{80} \left ( 2 \chat{A2} + 3 \chat{B2} \right ) \label{eq:muints} \\
\musidones = \frac{9}{280} \left ( 4 \chat{A1} + 3 \chat{B1} \right ) \label{eq:musidones} \\
\musidtwos = \frac{9}{140} \left ( \chat{A1} - \chat{B1} \right ) \label{eq:musidtwos} \\
\mubulkones = \frac{9}{16} \chat{A2} \label{eq:mubulkones} \\
\mubulktwos = \frac{9}{40} \chat{A1} \label{eq:mubulktwos} \\
\muintm = \frac{15}{2} \dhat{K2} + 5 \dhat{L2} + \dhat{M2} \label{eq:muintm} \\
\musidonem = 2 \dhat{L1} + \frac{4}{7} \dhat{M1} \label{eq:musidonem} \\
\musidtwom = 2 \dhat{S1} - \frac{4}{3} \dhat{L1} - \frac{8}{21} \dhat{M1} \label{eq:musidtwom} \\
\mubulkonem = \frac{15}{2} \dhat{S2} \label{eq:mubulkonem} \\
\mubulktwom = \frac{3}{2} \dhat{K1} + 2 \dhat{L1} + \dhat{M1} \label{eq:mubulktwom} \\
\chat{Xi} = \frac{1}{a^5} \intr r^4 q \scali[i]{f}^* X dr \quad \text{for }i=1,2\text{ and }X=A,B \label{eq:chat_def} \\
\dhat{Xi} = \frac{1}{a^3} \intr r^2 q \scali[i]{f}^* X dr \quad \text{for }i=1,2\text{ and }X=L,K,M,S \label{eq:dhat_def}
\end{gather}%

\vspace{1em}
\emph{Mobility and resistance coefficients:}
\begin{minipage}{0.5\textwidth}
\begin{align}
A = \frac{2}{r} \left ( x_{11}^g + x_{21}^g\right ) \label{eq:A_def} \\
B = \frac{4}{r} \left ( y_{11}^g + y_{21}^g\right ) \label{eq:B_def} \\
K = \frac{3}{20 \pi a^3} z_{1}^m - 1 \label{eq:K_def} \\
L = \frac{3}{20 \pi a^3} \left ( y_{1}^m - z_{1}^m \right ) \label{eq:L_def}
\end{align}
\end{minipage}%
\begin{minipage}{0.5\textwidth}
\begin{align}
P = \frac{r}{10 a} \left ( X_{11}^P - X_{12}^P \right ) \label{eq:P_def} \\
Q = \frac{2}{5} \left ( X_{11}^Q + X_{12}^Q \right ) \label{eq:Q_def} \\
S = Q - PA \label{eq:S_def} \\
N = S - \frac{2}{3}L-\frac{1}{3}M \label{eq:N_def}
\end{align}
\end{minipage}
\begin{equation}
M = \frac{3}{20 \pi a^3} \left ( \frac{3}{2} x_{1}^m - 2 y_{1}^m + \frac{1}{2} z_{1}^m \right ) \label{eq:M_def}
\end{equation}
\end{minipage}
}
\caption{Summary of viscosities, mobility and resistance coefficients required for the evaluation of the interaction stresses, defined in terms of secondary mobility and resistance coefficients taken from \citet{kim91} for $x_{11}^g$, $y_{12}^g$, $x_{1}^m$, $y_{1}^m$ and $z_{1}^m$ and \citet{jeffrey93} for $X_{11}^P$, $X_{12}^P$, $X_{11}^Q$ and $X_{12}^Q$ \label{fig:interaction_coefficients}}
\end{figure}

\section{Order of magnitude analysis and final averaged suspension equations \label{sec:resulting_equations}}

With all terms required for the momentum equations defined, the objective is now to simplify the resulting equations by using an order of magnitude analysis to determine what terms are relatively small and can be neglected.

Note that following equation (\ref{eq:fjf_parts}), $\vecti[j,\text{m}]{f}$ is $\order{a/L}$ relative to $\vecti[j,\text{s}]{f}$ so that both the fixed and interaction stresses acting on the mixture are in general smaller than those acting on the individual phases.  Hence, consistent with \citet{jackson97}'s dilute system analysis, justification for the neglect of a specific term from any of the momentum equations needs to consider its importance within the mixture momentum equation, as well as the individual phase equations.  This is accomplished in the remainder of this section.  In Appendix (\ref{sec:oom_fluidsolid}) we repeat this analysis for the fluid and solid phase momentum equations, finding that any terms that can be neglected from the mixture momentum equation can also be neglected from those equations as well.

Substituting equations (\ref{eq:totf}), (\ref{eq:fjf_parts}), (\ref{eq:ffixm_def}) and (\ref{eq:fintm_def2}) into equation (\ref{eq:momentumm1}) produces the complete mixture momentum equation
\begin{multline}
\orderunder{\frac{\partial \rhom \uvectrho }{\partial t}  + \vnabla \cdot \rhom \uvectrho \uvectrho}{\nicefrac{\rho\uvect^2}{L}} = 
- \orderunder{\vnabla \pf}{\nicefrac{\pf}{L}}
%
+ \orderunder{\vect[rst,f]{f}}{\nicefrac{\rho\ave[f]{\uf' \uf'}}{L}}
%
+ \orderunder{\vect[rst,s]{f}}{\nicefrac{\rhos\phis\ave[s]{\us' \us'}}{L}}
\\
+ \orderunder{\vecti[\rho]{f}}{\nicefrac{\rho\phis\uslip^2}{L}}
%
+ \orderunder{\vect[rot,m]{f}}{\nicefrac{\muf\phis\omegaslip}{L}}
- \orderunder{\vnabla \cdot \taudilm}{\nicefrac{\muf\uvect}{L^2}}
- \orderunder{\vnabla \cdot \tauintm}{\nicefrac{\muf\phis^2\uvect}{L^2}}
\\
- \orderunder{\vnabla \cdot \tens[slip,m]{\tau}}{\nicefrac{\muf\phis\uslip}{L^2}}
+ \orderunder{\rhom \g}{\rho \g}
+ \orderunder{\epsf}{\nicefrac{\rho \uvect^2 l}{L^2}}
+ \orderunder{\epss}{\nicefrac{\rho \phis \uvect^2 a}{L^2},\nicefrac{\rho \phis \g a^2}{L^2}}
\label{eq:momentumm3}
\end{multline}
where a `dilute' mixture phase stress $\taudilm$ has been defined as per equation (\ref{eq:taudilm_def}).  Order of magnitude estimates have been added to each term in equation (\ref{eq:momentumm3}) using the following approximations:  $\order{\uf}=\order{\us}=\order{\uvect}$, $\order{\lambdafix}=\order{\lambdaint}=1$, $\order{\rhos}=\order{\rhof}=\order{\rho}$ and $\order{\phif}=1$.  Additionally, we have defined $\uslip=\us-\uf$ as the linear slip velocity of the particles and $\omegaslip=\omegas-\omegaf$ as the angular slip velocity of the particles.

Comparing magnitudes in equation (\ref{eq:momentumm3}) shows that several terms can be neglected.  The averaging errors $\epsf$ and $\epss$ are either $l/L$, $a/L$ or $a^2/L^2$ smaller than other terms so can be omitted.  The rotation term $\vect[rot,m]{f}$ requires a magnitude for $\omegaslip$.  To find this we use the solid phase angular momentum equation derived by \citet{jackson97} for dilute systems,
\begin{equation}
\orderunder{\rhos \left [ \frac{\partial \omegas }{\partial t} + \us \cdot \vnabla \omegas \right ]}{\nicefrac{\rho\uvect\omegas}{L}}
= -\frac{15\muf}{a^2}\omegaslip
\label{eq:omegas_eqn}
\end{equation}
that shows that the solid angular velocity is continuously driven towards the fluid angular velocity.  Noting equation (\ref{eq:farfieldv1}) then leads to $\omegaslip=\order{\rho\uvect^2a^2/(\muf L^2)}$, and $\vect[rot,m]{f}=\order{\rho\uvect^2a^2/L^3}$.  This rotation term is $\order{a^2/L^2}$ smaller than the main inertial terms, and can be neglected.



For $\vnabla \cdot \tens[slip,m]{\tau}$ and $\vecti[\rho]{f}$ we need a magnitude for $\uslip$.  Combining equations (\ref{eq:momentumf3}) and (\ref{eq:momentums3}) such that the pressure gradient is removed, and noting equation (\ref{eq:fdrags}), we find
\begin{multline}
\vect[drag,s]{f} = -\frac{9}{2a^2} \phis\muf\lambdafix\uslip =
\orderunder{\phis \vnabla \cdot \taudilm - \vect[faxen,s]{f}}{\nicefrac{\muf\phis\uvect}{L^2}}
+ \orderunder{\phis \phif (\rhof-\rhos) \g}{\Delta \rho \phis \g} \\
\orderunder{\rhos \phif \left [ \frac{\partial \phis \us }{\partial t}  + \vnabla \cdot \phis \us \us \right ] - \rhof \phis \left [ \frac{\partial \phif \uf }{\partial t}  + \vnabla \cdot \phif \uf \uf \right ] }{\nicefrac{\rho\phis\uvect^2}{L}}
%
\label{eq:slipeq1}
\end{multline}
where $\Delta \rho=\rhof - \rhos$ and the Reynolds stress and smaller terms have been neglected.  Utilising previous order assumptions leads to
\begin{equation}
\uslip=\order{\frac{\uvect a^2}{L^2},\frac{\Delta \rho \g a^2}{\muf},\frac{\rho \uvect^2 a^2}{\muf L}}
\label{eq:slipeq2}
\end{equation}
which shows that $\vnabla \cdot \tens[slip,m]{\tau}$ is $\order{a^2/L^2}$ smaller than each comparable term in (\ref{eq:momentumm3}) and can be neglected.  The order of $\vecti[\rho]{f}$ is also expressed in terms of the magnitude for $\uslip$ in equation (\ref{eq:momentumm3}).  Utilising the above yields
\begin{equation}
\vecti[\rho]{f}=\phis\order{\dnum[a]{Ar}\frac{a}{L},\dnum[a]{Re}^2\frac{a^2}{L^2}}
\label{eq:frho}
\end{equation}
where $\dnum[a]{Ar}=\Delta \rho^2 g a^3/\muf^2$ and $\dnum[a]{Re}=\rho |\uvect|a/\muf$ are types of particle Archimedes and Reynolds numbers, respectively, that tend not to be large in the considered systems.  Hence, provided the conditions of equation (\ref{eq:frho}) are satisfied then $\vecti[\rho]{f}$ can be neglected from the mixture momentum equation.  (Note that if $\vecti[\rho]{f}$ is not negligible, as an alternative the L.H.S. of the mixture momentum equation can remain as the sum of the total derivatives from the fluid and solid momentum equations).

Finally, the Reynolds stress terms require magnitudes for $\ave[f]{\uf' \uf'}$ and $\ave[s]{\us' \us'}$.  These terms capture the transport of momentum at the macroscopic scale due to local fluid disturbances occurring around the particles, and fluctuations in particle velocities, respectively.  For a turbulent flow both terms may be significant.  However for a laminar flow in a semi-dilute suspension as targeted in this study, the magnitude of the $\uf'$ fluctuations either results from strain occurring around each particle or slip between the solid and fluid phases, and hence is of $\order{a\tensgammadotf,\uslip}=\order{a\uvect/L,\uslip}$.  Noting further that these fluctuations only occur within the vicinity of the particles we find $\ave[f]{\uf' \uf'}=\order{\phis \uf'^2}=\phis\order{\uvect^2a^2/L^2,\dnum[a]{Ar} g a,\dnum[a]{Re}\uvect^2a^2/L^2}$, or
\begin{equation}
\vect[rst,f]{f} = \order{\frac{\rho\phis\uvect^2a^2}{L^3},\dnum[a]{Ar}\frac{\rho\phis\g a}{L},\dnum[a]{Re}\frac{\rho\phis\uvect^2a^2}{L^3}}
\label{eq:frstf_mag}
\end{equation}
Comparing this expression with other terms in the mixture momentum balance shows that (conservatively) $\vect[rst,f]{f}$ can be neglected provided that $\dnum[a]{Ar}a/L \ll 1$ and $\dnum[a]{Re} a^2/L^2 \ll 1$.  Given that both $\dnum[a]{Ar}$ and $\dnum[a]{Re}$ tend not to be large for the solid-liquid suspensions targeted in this study, $\vect[rst,f]{f}$ can therefore generally be neglected.  Turning to the evaluation of $\vect[rst,s]{f}$, fluctuations in particle velocities occur due interactions with other particles during laminar suspension flow, so $\us'=\order{a\tensgammadots}=\order{a\uvect/L}$.  Noting that the frequency that a particular particle interacts within another particle scales with $\phis$ leads to $\ave[s]{\us' \us'}=\order{\phis \uvect^2 a^2 / L^2}$, or
\begin{equation}
\vect[rst,s]{f} = \order{\frac{\rho\phis\uvect^2a^2}{L^3}}
\label{eq:frsts_mag}
\end{equation}
As this magnitude is $\order{a^2/L^2}$ less than the inertial terms in equation (\ref{eq:momentumm3}), $\vect[rst,s]{f}$ can also be safely neglected under the considered circumstances.

In summary we have shown that $\epsf$, $\epss$, $\vect[rot,m]{f}$, $\vnabla \cdot \tens[slip,m]{\tau}$, $\vecti[\rho]{f}$, $\vect[rst,f]{f}$ and $\vect[rst,s]{f}$ are all less significant than other terms occurring in the mixture momentum equation (\ref{eq:momentumm3}), and can hence be neglected.  Further, analysis of the fluid and solid momentum equations (\ref{eq:momentumf3}) and (\ref{eq:momentums3}) as presented in Appendix \ref{sec:oom_fluidsolid} shows that these same terms can also be neglected from these individual phase momentum equations.  Hence, the final momentum equations resulting from the averaging analysis are given by equations (\ref{eq:momentumf2}) to (\ref{eq:momentumm2}) in the equation summary Figure \ref{fig:summary_equations}.

\section{Conclusions and future work}

Figures \ref{fig:summary_equations} and \ref{fig:interaction_coefficients} summarise the findings of this work.  Three simplified momentum equations are supplemented by equations for the interphase forces $\vect[drag,s]{f}$ and $\vect[faxen,s]{f}$, viscous stress definitions for the mixture ($\taudilm$ and $\tauintm$) and solid ($\tauints$) phases, definitions of strain rates that recognise the physical dimensions of the problem and potentially divergent nature of the phase velocities, and suspension viscosities that quantitatively link the micro-structure of the suspension to its bulk stresses.  In paper II we link this microstructure to particle roughness\citep{noori24a} and examine the subsequent performance of the equations for predicting particle migration in dilute and semi-dilute pressure driven suspension flows.

Linking macroscopic suspension properties such as pressure and strain-rates to microscopic analyses involves choices via the design of the closure models.  The stress closure model that has split the total stress on each particle as being from a fixed bed and from particle interactions has resulted in the non-Newtonian suspension scale interaction stresses being different for each phase, and the generalised Newtonian suspension scale fixed stresses acting only on the fluid phase.  This is consistent with the updates to the SB model\citep{nott11}, also based on a \citet{jackson97} type analysis.  The pressure gradient appears as acting proportionally on both phases in our equations, consistent with \citet{jackson98}'s corrected analysis and \citet{zhang94}'s analysis, but in difference to \citet{nott11}'s results.  This proportionality is a direct result of accounting for pressure in the closure model for the local stress surrounding each particle in the former studies.  \citet{nott11}'s dilute stress closure model only considers the viscous stresses acting on each particle, which appears to be an omission.

A key assumption in the \citet{batchelor72d} type analyses is that interacting particles are modelled as being force free during their binary interactions.  If applied through our analysis this assumption would not affect the mixture momentum equations (and subsequent suspension viscosity), but would result in no interaction stress being applied to the solid phase, with the potential effect of causing reverse shear-induced particle migration.  Our `semi-free' interaction model recognises that the modelled binary interactions between particles in a suspension can result in non-zero forces on particles, as each particle is interacting with more than one particle at any one time, and it is the sum of these forces that results in the suspension level stress.

The interpretation of $\lambdafix$ as being a stress multiplier around the surface of all of the particles due to particle crowding (i.e., multi-particle effects), has resulted in a faxen force that involves gradients of this parameter.  The form of this force needs validation.  Via this closure parameter, the averaging process has also linked the suspension's hindered settling function to its dilute shear viscosity, being two well studied functions.  The validity of this result is examined in paper II.  $\lambdaint$ conceptually models the same increase in stress as $\lambdafix$, but during the microhydrodynamic particle interactions.  On this basis then $\lambdaint$ could justifiably be set equal to $\lambdafix$.  $\lambdaint$ is more important at higher volume fractions.

There are many aspects of this work that could be improved.  Contact forces are neglected, except for the effect they have on the suspension microstructure.  This is justified given the scope of the work (semi-dilute), but note that the normal contact forces would be well modelled by the normal hydrodynamic forces employed in this model regardless, as these are determined by the drag on each sphere when moving as a pair.  Tangential or friction forces may not be well modelled by the hydrodynamic interaction however, but could be incorporated into the model via alternative mobility functions.  Attempts have been made to model the stresses due to bed dilation or compaction (`bulk' viscosities), however this has been viewed as secondary to modelling deviatoric strain:  Specifically, because the structure of the microstructure PDF is only a function of the deviatoric field, and not the rate of dilation, the resulting model is best viewed as applicable to flow fields that involve small amounts of dilational strain relative to the deviatoric strain.  This, however, is probably a reasonable assumption for most application systems, but again could form an avenue for future work.  Indeed, a strength of the present analysis is that it links suspension microstructure to bulk behaviour using a defined set of closure assumptions, allowing future models for the microstructure (described by the $\scali[i]{f}^*$ parameters) to be easily converted to continuum-based suspension-scale constitutive equations.

\backsection[Acknowledgements]{The author greatly appreciated discussions with Prof Krishnaswamy Nandakumar during 2001--03 on this topic, and with Dr Mohammad Noori, Dr Nilanka Ekanayake and Dr Joseph Berry during the period 2017--24 on using these equations to simulate suspension flows.}

\backsection[Funding]{The author gratefully acknowledges the support of Alberta Ingenuity, Canada, in providing an Ingenuity Associateship (2002--03) that supported the early development of this theory; `A multi-fluid description of solid liquid suspension flows'.}

\backsection[Declaration of interests]{The authors report no conflict of interest.}

\backsection[Author ORCIDs]{D.J.E. Harvie, https://orcid.org/0000-0002-8501-1344}

\appendix

\section{Appendix}

\subsection{Averaging identities \label{sec:averaging_identities}}

Here we derive averaging identities that follow from the fluid, solid and particle averaging definitions and that are used in the main analysis.

\subsubsection{Repeated averaging \label{sec:averaging_repeated}}

Starting with the fluid based average of two generic variables $f_1$ and $f_2$, we wish to find an approximation to the following expression in terms of only individually averaged variables:
\begin{equation}
\phif(\x) \ave[f]{\ave[f]{f_1}f_2}(\x) = \int_{\Vf} \ave[f]{f_1}(\y) f_2(\y) g(|\x-\y|) d\scali[\y]{V} \label{eq:aveidf1f2_1}
\end{equation}
Expanding $\ave[f]{f_1}(\y)$ around $\x$ using a Taylor series gives
\begin{equation}
\ave[f]{f_1}(\y) = \ave[f]{f_1}(\x) + (\y-\x) \cdot \vnabla[\x] \ave[f]{f_1}(\x) + \order{\frac{{|\y-\x|}^2 f_1}{L^2}} \nonumber
\end{equation}
noting that the fluid averaged variable $\ave[f]{f_1}$ varies over the macroscopic lengthscale $L$.  Substituting this series into equation (\ref{eq:aveidf1f2_1}) leads to
\begin{align}
\phif(\x) \ave[f]{\ave[f]{f_1}f_2}(\x) & = \phif(\x) \ave[f]{f_1}(\x) \ave[f]{f_2}(\x) \nonumber \\
& + \vnabla[\x] \ave[f]{f_1}(\x) \cdot \int_{\Vf} (\y-\x) f_2(\y) g(|\x-\y|) d\scali[\y]{V} \nonumber
\end{align}
where terms of order $l/L$ relative to those remaining have been neglected.  Recognising that the integral in this equation is $\order{l f_2 \phif}$ we find that
\begin{equation}
\ave{\ave{f_1}f_2} = \ave{f_1}\ave{f_2} + \order{\frac{l f_1 f_2}{L}} \label{eq:aveidf1f2_2}
\end{equation}
This analysis applies analogously to repeated application of solid phase averages ($\ave[s]{}$) and with minor modification to particle phase averages ($\ave[p]{}$) as well.  For this reason the f subscript has been dropped from the averaging operators in equation (\ref{eq:aveidf1f2_2}) as the identity applies to all averaging types used (i.e., fluid, solid, mixture and particle).  Hence, provided the separation of lengthscales $l \ll L$ constraint is satisfied any averaged variable can be extracted from an averaged group of variables with $\order{l/L}$ error, relative to the remaining terms.

Equation (\ref{eq:aveidf1f2_2}) can also be used to derive other repeated averaging identities.  Substituting $f_2=1$ in equation (\ref{eq:aveidf1f2_2}), or alternatively replacing $f_2$ by an averaged variable in the analysis leads to
\begin{equation}
\ave{\ave{f_1}} = \ave{f_1} + \order{\frac{l f_1}{L}} \label{eq:aveidaf1}
\end{equation}
and
\begin{equation}
\ave{\ave{f_1}\ave{f_2}} = \ave{f_1}\ave{f_2} + \order{\frac{l f_1 f_2}{L}} \label{eq:aveidaf1af2}
\end{equation}
respectively, showing again that neglecting repeated averaging operations results in an error of order $\order{l/L}$ relative to the remaining terms.  Again, both equations (\ref{eq:aveidaf1}) and (\ref{eq:aveidaf1af2}) apply to all averaging types (fluid, solid, particle and mixture).

We note that \citet{jackson97} uses a less conservative assumption that replacing the average of an averaged variable with the average variable (that is, equation (\ref{eq:aveidaf1})) results in a relative $\order{l^2/L^2}$ error.  We are unable to reproduce the analysis that suggests this, but given the separation of lengthscales this does not affect subsequent results in this paper.

\subsubsection{Equivalence of solid and particle phase averages \label{sec:averaging_solid_particle}}

Here we quantify the error in replacing particle averaged variables by their solid averaged equivalents.  Starting from the solid phase average definition
\begin{align}
\phis(\x) \ave[s]{f}(\x) &= \int_{\Vs} g(|\x-\y|) f(\y) d\scali[\y]{V} \nonumber \\
 &= \sum_p \int_{\Vp} g(|\x-\y|) f(\y) d\scali[\y]{V} \nonumber
\end{align}
we use a Taylor series expansion of $g(\x-\y|)$ around $\xp$ to give
\begin{align}
\phis(\x) \ave[s]{f}(\x) &= \sum_p \int_{\Vp} g(|\x-\xp|) f(\y) d\scali[\y]{V} \nonumber \\
& +  \sum_p \int_{\Vp} \vnablaxp g(|\x-\xp|) \cdot (\y-\xp) f(\y) d\scali[\y]{V} \nonumber \\
& +  \sum_p \int_{\Vp} \frac{1}{2} \vnablaxp \vnablaxp g(|\x-\xp|) : (\y-\xp)^2 f(\y) d\scali[\y]{V} + \ldots \nonumber
\end{align}
Recognising via the chain rule that $\vnablaxp^n g(|\x-\xp|) = (-1)^n \vnablax^n g(|\x-\xp|)$ and that terms independent of $\y$ can be removed from the integrals we find
\begin{align}
\phis(\x) \ave[s]{f}(\x) &= \sum_p g(|\x-\xp|) \int_{\Vp} f(\y) d\scali[\y]{V} \nonumber \\
                         &- \vnablax \cdot \sum_p g(|\x-\xp|) \int_{\Vp} (\y-\xp) f(\y) d\scali[\y]{V} \nonumber \\
                         &+ \vnablax \vnablax : \sum_p g(|\x-\xp|) \frac{1}{2} \int_{\Vp} (\y-\xp)^2 f(\y) d\scali[\y]{V} - \ldots
\label{eq:aveidsolpart_2}
\end{align}
showing that a solid averaged property can be expressed as a series of particle averaged property moments.

We now apply equation (\ref{eq:aveidsolpart_2}) to specific particle properties.  For the volume fraction equivalent we substitute $f=1$ into the equation giving
\begin{equation}
\phis = n \nu + \order{\phis \frac{a^2}{L^2}}
\label{eq:aveidphis}
\end{equation}
where the first, second and third integrals in equation (\ref{eq:aveidsolpart_2}) are evaluated as $\nu$, zero (due to the definition of $\xp = (1/\nu) \int_{\Vp} \y d\scali[\y]{V}$), and $\order{a^2 \nu}$, respectively.

For velocity we substitute $f=\vect{u}$ into equation (\ref{eq:aveidsolpart_2}) giving
\begin{align}
\phis(\x) \ave[s]{\vect{u}}(\x) &= \sum_p g(|\x-\xp|) \int_{\Vp} \vect{u}(\y) d\scali[\y]{V} \nonumber \\
                         &- \vnablax \cdot \sum_p g(|\x-\xp|) \int_{\Vp} (\y-\xp) \vect{u}(\y) d\scali[\y]{V} \nonumber \\
                         &+ \vnablax \vnablax : \sum_p g(|\x-\xp|) \frac{1}{2} \int_{\Vp} (\y-\xp)^2 \vect{u}(\y) d\scali[\y]{V} - \ldots
\label{eq:aveidus_1}
\end{align}
Recognising that within each particle $p$ the solid body motion can be expressed using the particle's velocity $\up$ and angular velocity $\omegap$ as
\begin{equation}
\vect{u}(\y) = \up + \omegap \cross (\y-\xp) \nonumber
\end{equation}
the three integrals in equation (\ref{eq:aveidus_1}) can be evaluated as $\nu \up$, $\order{a^2 \nu \omegas}$ and $\order{\nu a^2 \us}+\order{\nu a^3 \omegas}$, respectively.  Here $\omegas$ is the solid phase averaged angular velocity of the particles.  There are two possible assumptions for evaluating the magnitude of $\omegas$.  As shown by \citet{jackson97} and discussed in section \ref{sec:resulting_equations}, under dilute conditions this angular velocity is continuously driven towards the angular velocity of the fluid, leading to $\order{\omegas}=\order{\us/L}$.  However under concentrated conditions solid particles can `jam' or alternatively slip past each other along fault lines, suggesting instead a higher estimate of $\order{\omegas}=\order{\us/a}$.  For the semi-dilute conditions targeted in this study the lower estimate for $\order{\omegas}$ is physically more appropriate, however even using the high concentration estimate we find
\begin{equation}
\phis \ave[s]{\vect{u}} = n \nu \ave[p]{\vect{u}} + \order{\phis \frac{a}{L} \us}
\label{eq:aveidus}
\end{equation}
showing that solid and particle averaged velocities can be interchanged, given the separation of lengthscales.

\subsubsection{Reynolds stress averages \label{sec:averaging_rst}}

In this section we decompose the inertial advection terms, thus defining the Reynolds stress terms and evaluating the error associated with this averaging process.

Commencing with the fluid phase, the average of the fluid velocity dyadic can be written as
\begin{equation}
\ave[f]{\vect{u}\vect{u}} =  \ave[f]{\uf \uf} + \ave[f]{\uf \uf'} + \ave[f]{\uf' \uf} + \ave[f]{\uf' \uf'} \nonumber
\label{eq:aveiduuf_1}
\end{equation}
where as per the main text, the perturbation velocity of the fluid phase is defined within the fluid as $\uf'=\vect{u}-\uf$.  Now, utilising the repeated average equations (\ref{eq:aveidf1f2_2}) and (\ref{eq:aveidaf1af2}) we find
\begin{equation}
\ave[f]{\vect{u}\vect{u}} =  \uf \uf + \uf \ave[f]{\uf'} + \ave[f]{\uf'}\uf + \ave[f]{\uf' \uf'} + \order{\frac{l}{L} \uf^2}
\label{eq:aveiduuf_2}
\end{equation}
however
\begin{equation}
\ave[f]{\uf'} = \ave[f]{\vect{u}-\uf} = \ave[f]{\vect{u}}-\ave[f]{\uf} = \uf - \uf + \order{\frac{l}{L} \uf} = \order{\frac{l}{L} \uf} \nonumber
\label{eq:aveiduuf_3}
\end{equation}
Therefore the two averaged perturbation terms can be removed from equation (\ref{eq:aveiduuf_2}) with no increase in error, giving finally the following relationship for the fluid phase inertial decomposition,
\begin{equation}
\phif \ave[f]{\vect{u}\vect{u}} = \phif \uf \uf + \phif \ave[f]{\uf' \uf'} + \order{\phif \frac{l}{L} \uf^2}
\label{eq:aveiduuf}
\end{equation}

The solid phase analysis is slightly more complex as we wish to express the Reynolds stress term using solid phase averaged variables, rather than particle phase ones.  Recognising that the repeated averaging identities used in the above fluid phase analysis apply also to the particle phase, the solid phase inertial advection term can be written
\begin{equation}
n \nu \ave[p]{\vect{u}\vect{u}} = n \nu \upp \upp + n \nu \ave[p]{\upp' \upp'} + \order{\phis \frac{l}{L} \us^2}
\label{eq:aveiduus_1}
\end{equation}
where the particle based perturbation velocity is defined as $\upp'^p = \up - \upp(\xp)$.  Now, using equations (\ref{eq:aveidphis}) and (\ref{eq:aveidus}) the first term on the right can be written
\begin{align}
n \nu \upp \upp & = \left [ \phis \us +  \order{\phis \frac{a}{L} \us} \right ] \left [ \left ( 1 + \order{\frac{a^2}{L^2}} \right ) \us +  \order{ \frac{a}{L} \us} \right ] \nonumber \\
& = \phis \us \us + \order{\phis \frac{a}{L} \us^2}
\label{eq:aveiduus_2}
\end{align}
For the second term we first define the solid based perturbation velocity within the volume of each particle $p$ as $\us'(\y) = \up - \us(\y)$, and then expand the following inertial term using Taylor series expansions of both $g(\x-\y|)$ and $\us(\y)$ around $\xp$ to give
\begin{align}
\phis \ave[s]{\us' \us'} & = \sum_p \int_{\Vp} g(|\x-\y|) ( \up - \us(\y) )  ( \up - \us(\y) ) d\scali[\y]{V} \nonumber \\
& = \sum_p g(|\x-\xp|) \int_{\Vp} ( \up - \us(\y) )  ( \up - \us(\y) ) d\scali[\y]{V} \nonumber \\
                         &- \vnablax \cdot \sum_p g(|\x-\xp|) \int_{\Vp} (\y-\xp) ( \up - \us(\y) )  ( \up - \us(\y) ) d\scali[\y]{V} \nonumber \\
& = \sum_p g(|\x-\xp|) \int_{\Vp} \left [ \up - \us(\xp) + \order{(\y-\xp)\us} \right ] \times \nonumber \\
& \quad \quad \left [ \up - \us(\xp) + \order{(\y-\xp)\us} \right ] d\scali[\y]{V} + \order{n \nu \frac{a}{L} \us^2} \nonumber
\end{align}
Evaluating the all the terms within the integral that involve the truncation error $\order{(\y-\xp)\us}$, we find that overall they contribute the same error to the above expression as the final error that is already included in the expression.  So recognising that $\up - \us(\xp)$ does not depend on $\y$ we arrive at
\begin{align}
\phis \ave[s]{\us' \us'} & = \nu \sum_p g(|\x-\xp|) \left [ \up - \us(\xp) \right ] 
\left [ \up - \us(\xp) \right ] + \order{n \nu \frac{a}{L} \us^2} \nonumber \\
& = n \nu \ave[p]{\upp' \upp'} + \order{\phis \frac{a}{L} \us^2}
\label{eq:aveiduus_4}
\end{align}
Finally substituting results from equations (\ref{eq:aveiduus_2}) and (\ref{eq:aveiduus_4}) into equation (\ref{eq:aveiduus_1}) gives the final identity
\begin{equation}
n \nu \ave[p]{\vect{u}\vect{u}} = \phis \us \us + \phis \ave[s]{\us' \us'} + \order{\phis \frac{a}{L} \us^2} 
\label{eq:aveiduup}
\end{equation}
Note that if we had instead defined the solid perturbation velocity using the local solid velocity at $\y$ rather than the particle velocity $\up$ the error in equation (\ref{eq:aveiduup}) using our high concentration assumption for $\omegas$ would have been $\order{\phis \us^2}$, and hence non-negligible.  This subtle difference highlights the care required in evaluating the solid phase Reynolds stress term to ensure non-negligible averaging errors.

\subsection{Dimensional operator notation\label{sec:dimensional_operator}}

The dimensional operator notation $[...]_\text{D}$ means to evaluate all of the contents of the brackets, recursively if necessarily, in only the physical dimensions of the problem, and to leave all other elements zero.  To represent this mathematically in a two dimensional problem using tensor index notation, used for example in the microstructure algebra, a dimensional delta function $d_i$ is included in every dimensional sum, with this delta defined as
\begin{equation}
d_i = 1 - \delta_{iD}
\label{eq:d_i}
\end{equation}
where $D$ is the number of the dimension that is not physical.  Note that if using implied Einstein summation convention the $d_i$ subscript does not imply summation.  In three dimensional problems $d_i=1$ always and hence the $[...]_\text{D}$ operator can be ignored.

To illustrate, the physical dimension unit tensor is
\begin{equation}
\ItensD = \left [ \Itens \right ]_\text{D} = \sum_{i,j}=\delta_{ij}d_i d_j\vecti[i]{\delta}\vecti[j]{\delta}
\label{eq:ItensD_index}
\end{equation}
while some strain rate related variables are,
\begin{equation}
\gammadotsph = \frac{2}{N} \left [ \vnabla \cdot \vect{u} \right ]_\text{D} = \frac{2}{N} \sum_{i} \frac{\partial u_i}{\partial x_i} d_i
\label{eq:gammasph_index}
\end{equation}
\begin{equation}
\tensgammadot = \left [ \vnabla \vect{u} - ( \vnabla \vect{u} )^\text{T} - \gammadotsph \Itens \right ]_\text{D} = \sum_{i,j} \left [ \frac{\partial u_j}{\partial x_i} + \frac{\partial u_i}{\partial x_j} - \gammadotsph \delta_{ij} \right ] d_i d_j\vecti[i]{\delta}\vecti[j]{\delta}
\label{eq:gammadot_index}
\end{equation}
\begin{align}
\Kp & = \frac{1}{2} \left [ \vnabla^2 \uf - \frac{N}{2(2+N)}  \sum_{i,j,k} \left ( \delta_{ij}\frac{\partial \gammadotsphf}{\partial x_k} + \delta_{ik}\frac{\partial \gammadotsphf}{\partial x_j} + \delta_{jk}\frac{\partial \gammadotsphf}{\partial x_i} \right )  \vecti[i]{\delta}\vecti[j]{\delta}\vecti[k]{\delta}  \right ]_\text{D} \nonumber \\
& = \frac{1}{2} \sum_{i,j,k} \left [ \frac{\partial^2 u_{\text{f},k}}{\partial x_i \partial x_j}  - \frac{N}{2(2+N)} \left ( \delta_{ij}\frac{\partial \gammadotsphf}{\partial x_k} + \delta_{ik}\frac{\partial \gammadotsphf}{\partial x_j} + \delta_{jk}\frac{\partial \gammadotsphf}{\partial x_i} \right )  \right ] d_i d_j d_j \vecti[i]{\delta}\vecti[j]{\delta}\vecti[k]{\delta}
\label{eq:Kp_index}
\end{align}
where $N$ is the number of physical dimensions for the problem.

\subsection{Spherical integral identities\label{sec:integrals}}

The following integrals are used in the derivation of the collision stress tensor.  They are evaluated around a sphere surface $S_r$ of radius $r$, assuming that $\tensgammadot$ is symmetric and traceless, and that $\n$ is an outward pointing normal to the sphere surface:
\begin{align}
\tensi[1]{I} & = \int_{S_r} \n\n \left ( \tensgammadot : \n\n \right ) dS = \frac{8\pi r^2}{15} \tensgammadot \label{eq:I1}\\
\tensi[2]{I} & = \int_{S_r} \n \left ( \tensgammadot \cdot \n \right ) dS = \int_{S_r} \left ( \tensgammadot \cdot \n \right ) \n dS \nonumber \\
& = \int_{S_r} \n \left ( \n \cdot \tensgammadot \right ) dS = \int_{S_r} \left ( \n \cdot \tensgammadot \right ) \n dS = \frac{4\pi r^2}{3} \tensgammadot \label{eq:I2}\\
\tensi[3]{I} & = \int_{S_r} \n\n \left ( \tensgammadot : \n\n \right ) \left ( \tensgammadot : \n\n \right ) dS = \frac{16\pi r^2}{105} \left [ 2 \tensgammadot \cdot \tensgammadot + 4 \Itens \right ] \label{eq:I3}\\
\tensi[4]{I} & = \int_{S_r} \n \left ( \tensgammadot \cdot \n \right ) \left ( \tensgammadot : \n\n \right ) dS = \int_{S_r} \left ( \tensgammadot \cdot \n \right ) \n \left ( \tensgammadot : \n\n \right ) dS \nonumber \\
& = \int_{S_r} \n \left ( \n \cdot \tensgammadot \right ) \left ( \tensgammadot : \n\n \right ) dS = \int_{S_r} \left ( \n \cdot \tensgammadot \right ) \n \left ( \tensgammadot : \n\n \right ) dS \nonumber \\
& = \frac{8\pi r^2}{15}  \tensgammadot \cdot \tensgammadot \label{eq:I4} \\
\tensi[5]{I} & = \int_{S_r} \n \n dS = \frac{4\pi r^2}{3} \Itens \label{eq:I5}\\
\scali[6]{I} & = \int_{S_r} dS = 4\pi r^2 \label{eq:I6}\\
\scali[7]{I} & = \int_{S_r} \tensgammadot : \n\n dS = 0 \label{eq:I7}\\
\scali[8]{I} & = \int_{S_r} \left ( \tensgammadot : \n\n \right ) \left ( \tensgammadot : \n\n \right ) dS = \frac{16\pi r^2}{15} \gammadot^2 \label{eq:I8}\\
\tensi[9]{I} & = \int_{S_r} \n \Itens \cdot \left( \vect{\Omega} \cross \n \right ) dS = \frac{4\pi r^2}{3} \epsilon_{lik}\Omega_l \vecti[i]{\delta}\vecti[k]{\delta} \label{eq:I9}\\
\tensi[10]{I} & = \int_{S_r} \n \n \n \cdot \left( \vect{\Omega} \cross \n \right ) dS = \tens{0} \label{eq:I10}
\end{align}
where $\gammadot = \sqrt{\frac{1}{2}\tensgammadot:\tensgammadot}$.  In deriving these relationships we have used the following identities:
\begin{align}
\int_{S_r} n_i n_j dS & = \frac{4 \pi r^2}{3} \delta_{ij} \label{eq:intrel1} \\
\int_{S_r} n_i n_j n_k n_l dS & = \frac{4 \pi r^2}{15} \left ( \delta_{ij}\delta_{kl} + \delta_{ik}\delta_{lj} +\delta_{il}\delta_{kj} \right ) \label{eq:intrel2} \\
\int_{S_r} n_i n_j n_k n_l n_m n_n dS & = \frac{4 \pi r^2}{105} \left [ 
\delta_{ij} \left ( \delta_{kl}\delta_{mn} + \delta_{km}\delta_{ln} +\delta_{kn}\delta_{lm} \right ) \right . \nonumber\\
& \quad + \delta_{ik} \left ( \delta_{jl}\delta_{mn} + \delta_{jm}\delta_{ln} +\delta_{jn}\delta_{lm} \right )\nonumber\\
& \quad + \delta_{il} \left ( \delta_{jk}\delta_{mn} + \delta_{jm}\delta_{kn} +\delta_{jn}\delta_{km} \right ) \nonumber\\
& \quad + \delta_{im} \left ( \delta_{jk}\delta_{ln} + \delta_{jl}\delta_{kn} +\delta_{jn}\delta_{kl} \right ) \nonumber\\
& \quad \left . + \delta_{in} \left ( \delta_{jk}\delta_{lm} + \delta_{jl}\delta_{km} +\delta_{jm}\delta_{kl} \right ) \right ]
\label{eq:intrel3}
\end{align}
The first two of these identities are quoted in \citet{jackson97}.  The third was calculated using computational symbolic algebra applied to all the possible combinations of the six normal coefficients.

\subsection{Order of magnitude analysis on fluid and solid momentum equations\label{sec:oom_fluidsolid}}

Substituting equations (\ref{eq:totf}), (\ref{eq:fjf_parts}), (\ref{eq:ffixs_def4}), (\ref{eq:ffixm_def}), (\ref{eq:fints_def}) and (\ref{eq:fintm_def2}) into equation (\ref{eq:momentumf1}) leads to the complete fluid phase momentum equation,
\begin{multline}
\orderunder{\rhof \left [ \frac{\partial \phif \uf }{\partial t}  + \vnabla \cdot \phif \uf \uf \right ]}{\nicefrac{\rho\uvect^2}{L}} = 
- \orderunder{\phif \vnabla \pf}{\nicefrac{\pf}{L}}
- \orderunder{\vect[drag,s]{f}}{\nicefrac{\muf\phis\uslip}{a^2}}
- \orderunder{\vect[faxen,s]{f}}{\nicefrac{\muf\phis\uvect}{L^2}}
\\
+ \orderunder{\vect[rst,f]{f}}{\nicefrac{\rho\phis\ave[f]{\uf' \uf'}}{L}}
%
+ \orderunder{\vect[rot,m]{f}}{\nicefrac{\muf\phis\omegaslip}{L}}
%
- \orderunder{\vect[rot,s]{f}}{\nicefrac{\muf\phis\omegaslip}{L}}
- \orderunder{\vnabla \cdot \tens[slip,m]{\tau}}{\nicefrac{\muf\phis\uslip}{L^2}} \\
- \orderunder{\vnabla \cdot \taudilm}{\nicefrac{\muf\uvect}{L^2}}
- \orderunder{\vnabla \cdot \tauintm}{\nicefrac{\muf\phis^2\uvect}{L^2}}
+ \orderunder{\vnabla \cdot \tauints}{\nicefrac{\muf\phis^2\uvect}{L^2}}
+ \orderunder{\rhof \phif \g}{\rho \g}
+ \orderunder{\epsf}{\nicefrac{\rho \uvect^2 l}{L^2}}
\label{eq:momentumf3}
\end{multline}
Similarly, substituting the same equations into equation (\ref{eq:momentums1}) leads to the complete solid phase momentum equation,
\begin{multline}
\orderunder{\rhos \left [ \frac{\partial \phis \us }{\partial t}  + \vnabla \cdot \phis \us \us \right ]}{\nicefrac{\rho\phis\uvect^2}{L}} = 
- \orderunder{\phis \vnabla \pf}{\nicefrac{\phis\pf}{L}}
+ \orderunder{\vect[drag,s]{f}}{\nicefrac{\muf\phis\uslip}{a^2}}
+ \orderunder{\vect[faxen,s]{f}}{\nicefrac{\muf\phis\uvect}{L^2}}
\\
+ \orderunder{\vect[rst,s]{f}}{\nicefrac{\rho\phis\ave[s]{\us' \us'}}{L}}
+ \orderunder{\vect[rot,s]{f}}{\nicefrac{\muf\phis\omegaslip}{L}}
- \orderunder{\vnabla \cdot \tauints}{\nicefrac{\muf\phis^2\uvect}{L^2}}
+ \orderunder{\rhos \phis \g}{\rho \phis \g} 
+ \orderunder{\epss}{\nicefrac{\rho \phis \uvect^2 a}{L^2},\nicefrac{\rho \phis \g a^2}{L^2}}
\label{eq:momentums3}
\end{multline}
Utilising the order of magnitude estimates developed in section \ref{sec:resulting_equations} we see that $\epsf$, $\epss$, $\vect[rot,m]{f}$, $\vnabla \cdot \tens[slip,m]{\tau}$, $\vecti[\rho]{f}$, $\vect[rst,f]{f}$ and $\vect[rst,s]{f}$ are each included or have equivalently sized terms in the mixture momentum equation (\ref{eq:momentumm3}). So by analogy with the analysis of that section, all of these terms can be neglected from the above individual phase momentum equations and the final phase-specific momentum equations (\ref{eq:momentumf2}) and (\ref{eq:momentums2}) result.

\bibliography{abbreviated,daltonh,main}

\ifx\undefined\allcaps\def\allcaps#1{#1}\fi
\begin{thebibliography}{43}
\expandafter\ifx\csname natexlab\endcsname\relax\def\natexlab#1{#1}\fi
\def\au#1{#1} \def\ed#1{#1} \def\yr#1{#1}\def\at#1{#1}\def\jt#1{\textit{#1}}
  \def\bt#1{#1}\def\bvol#1{\textbf{#1}} \def\vol#1{#1} \def\pg#1{#1}
  \def\publ#1{#1}\def\arxiv#1{#1}\def\org#1{#1}\def\st#1{\textit{#1}}

\bibitem[Anderson \& Jackson(1967)]{anderson67}
{\sc \au{Anderson, T.~B.} \& \au{Jackson, Roy}} \yr{1967}  \at{Fluid
  {Mechanical} {Description} of {Fluidized} {Beds}. {Equations} of {Motion}}.
  \jt{Industrial \& Engineering Chemistry Fundamentals}  \bvol{6}~(4),
  \pg{527--539}.

\bibitem[Arp \& Mason(1977)]{arp77}
{\sc \au{Arp, P.~A} \& \au{Mason, S.~G}} \yr{1977}  \at{The kinetics of flowing
  dispersions: {VIII}. {Doublets} of rigid spheres (theoretical)}.  \jt{Journal
  of Colloid and Interface Science}  \bvol{61}~(1),  \pg{21--43}.

\bibitem[Batchelor \& Green(1972)]{batchelor72d}
{\sc \au{Batchelor, G.~K.} \& \au{Green, J.~T.}} \yr{1972}  \at{The
  determination of the bulk stress in a suspension of spherical particles to
  order c2}.  \jt{Journal of Fluid Mechanics}  \bvol{56}~(3),  \pg{401--427}.

\bibitem[Blanc(2012)]{blanc12a}
{\sc \au{Blanc, Fr{\'e}d{\'e}ric}} \yr{2012}  \at{Rh{\'e}ologie et
  microstructure des suspensions concentr{\'e}es non browniennes}. PhD thesis,
  Laboratoire de Physique de la Mati{\`e}re Condens{\'e}e.

\bibitem[Blanc {\em et~al.\/}(2012)Blanc, Lemaire, Meunier \& Peters]{blanc12}
{\sc \au{Blanc, Fr{\'e}d{\'e}ric}, \au{Lemaire, Elisabeth}, \au{Meunier, Alain}
  \& \au{Peters, Fran{\c c}ois}} \yr{2012}  \at{Microstructure in sheared
  non-{Brownian} concentrated suspensions}.  \jt{Journal of Rheology}
  \bvol{57}~(1),  \pg{273--292}.

\bibitem[Blanc {\em et~al.\/}(2011)Blanc, Peters \& Lemaire]{blanc11b}
{\sc \au{Blanc, Fr{\'e}d{\'e}ric}, \au{Peters, Fran{\c c}ois} \& \au{Lemaire,
  Elisabeth}} \yr{2011}  \at{Experimental {Signature} of the {Pair}
  {Trajectories} of {Rough} {Spheres} in the {Shear}-{Induced} {Microstructure}
  in {Noncolloidal} {Suspensions}}.  \jt{Physical Review Letters}
  \bvol{107}~(20),  \pg{208302}.

\bibitem[Brady {\em et~al.\/}(2006)Brady, Khair \& Swaroop]{brady06}
{\sc \au{Brady, John~F.}, \au{Khair, Aditya~S.} \& \au{Swaroop, Manuj}}
  \yr{2006}  \at{On the bulk viscosity of suspensions}.  \jt{Journal of Fluid
  Mechanics}  \bvol{554},  \pg{109--123}.

\bibitem[Brady \& Morris(1997)]{brady97}
{\sc \au{Brady, John~F.} \& \au{Morris, Jeffrey~F.}} \yr{1997}
  \at{Microstructure of strongly sheared suspensions and its impact on rheology
  and diffusion}.  \jt{J. Fluid Mech.}  \bvol{348},  \pg{103--139}.

\bibitem[Da~Cunha \& Hinch(1996)]{dacunha96}
{\sc \au{Da~Cunha, F.~R.} \& \au{Hinch, E.~J.}} \yr{1996}  \at{Shear-induced
  dispersion in a dilute suspension of rough spheres}.  \jt{Journal of Fluid
  Mechanics}  \bvol{309},  \pg{211--223}.

\bibitem[Drazer {\em et~al.\/}(2004)Drazer, Koplik, Khusid \&
  Acrivos]{drazer04}
{\sc \au{Drazer, German}, \au{Koplik, Joel}, \au{Khusid, Boris} \& \au{Acrivos,
  Andreas}} \yr{2004}  \at{Microstructure and velocity fluctuations in sheared
  suspensions}.  \jt{Journal of Fluid Mechanics}  \bvol{511},  \pg{237--263}.

\bibitem[Drew(1976)]{drew76}
{\sc \au{Drew, Donald~A.}} \yr{1976}  \at{Two-phase flows: {Constitutive}
  equations for lift and {Brownian} motion and some basic flows}.  \jt{Archive
  for Rational Mechanics and Analysis}  \bvol{62}~(2),  \pg{149--163}.

\bibitem[Drew \& Mandyam(1998)]{drew98}
{\sc \au{Drew, D.~A.} \& \au{Mandyam, H.}} \yr{1998}  \at{Effective {Media}
  {Theory} {Using} {Nearest} {Neighbor} {Pair} {Distributions}}.  \bt{In {\em
  Particulate {Flows}: {Processing} and {Rheology}\/} (ed. \ed{Donald~A. Drew,
  Daniel~D. Joseph \& Stephen~L. Passman})},  \pg{pp. 23--53}.  \publ{New York,
  NY: Springer New York}.

\bibitem[Gao {\em et~al.\/}(2010)Gao, Kulkarni, Morris \& Gilchrist]{gao10}
{\sc \au{Gao, C.}, \au{Kulkarni, S.~D.}, \au{Morris, J.~F.} \& \au{Gilchrist,
  J.~F.}} \yr{2010}  \at{Direct investigation of anisotropic suspension
  structure in pressure-driven flow}.  \jt{Physical Review E}  \bvol{81}~(4),
  \pg{041403}.

\bibitem[Gillissen \& Wilson(2018)]{gillissen18}
{\sc \au{Gillissen, J. J.~J.} \& \au{Wilson, H.~J.}} \yr{2018}  \at{Modeling
  sphere suspension microstructure and stress}.  \jt{Physical Review E}
  \bvol{98}~(3).

\bibitem[Gillissen \& Wilson(2019{\natexlab{{\em a\/}}})]{gillissen19}
{\sc \au{Gillissen, J. J.~J.} \& \au{Wilson, H.~J.}} \yr{2019{\natexlab{{\em
  a\/}}}}  \at{Effect of normal contact forces on the stress in shear rate
  invariant particle suspensions}.  \jt{Physical Review Fluids}  \bvol{4}~(1).

\bibitem[Gillissen \& Wilson(2019{\natexlab{{\em b\/}}})]{gillissen19a}
{\sc \au{Gillissen, J. J.~J.} \& \au{Wilson, H.~J.}} \yr{2019{\natexlab{{\em
  b\/}}}}  \at{Taylor-{Couette} instability in sphere suspensions}.
  \jt{Physical Review Fluids}  \bvol{4}~(4).

\bibitem[Guazzelli \& Pouliquen(2018)]{guazzelli18}
{\sc \au{Guazzelli, {\'E}lisabeth} \& \au{Pouliquen, Olivier}} \yr{2018}
  \at{Rheology of dense granular suspensions}.  \jt{Journal of Fluid Mechanics}
   \bvol{852}.

\bibitem[Jackson(1997)]{jackson97}
{\sc \au{Jackson, R.}} \yr{1997}  \at{Locally averaged equations of motion for
  a mixture of identical spherical particles and a {Newtonian} fluid}.
  \jt{Chemical Engineering Science}  \bvol{52}~(15),  \pg{2457--2469}.

\bibitem[Jackson(1998)]{jackson98}
{\sc \au{Jackson, R.}} \yr{1998}  \at{Erratum: Locally averaged equations of
  motion for a mixture of identical spherical particles and a newtonian fluid}.
   \jt{Chem. Eng. Sci.}  \bvol{53}~(10),  \pg{1955}.

\bibitem[Jeffrey {\em et~al.\/}(1993)Jeffrey, Morris \& Brady]{jeffrey93}
{\sc \au{Jeffrey, D.~J.}, \au{Morris, J.~F.} \& \au{Brady, J.~F.}} \yr{1993}
  \at{The pressure moments for two rigid spheres in low‐{Reynolds}‐number
  flow}.  \jt{Physics of Fluids A: Fluid Dynamics}  \bvol{5}~(10),
  \pg{2317--2325}, publisher: American Institute of Physics.

\bibitem[Joseph \& Lundgren(1990)]{joseph90}
{\sc \au{Joseph, D.} \& \au{Lundgren, T.}} \yr{1990}  \at{Ensemble averaged and
  mixture theory equations for incompressible fluid-particle suspensions}.
  \jt{Int. J. Multiphase Flow}  \bvol{16}~(1),  \pg{35--42}.

\bibitem[Kim \& Karrila(1991)]{kim91}
{\sc \au{Kim, Sangtae} \& \au{Karrila, Seppo~J.}} \yr{1991} {\em
  Microhydrodynamics: Principles and selected applications\/}.
  \publ{Butterworth-Heinemann}.

\bibitem[Leal(1992)]{leal92}
{\sc \au{Leal, L.~Gary}} \yr{1992} {\em Laminar flow and convective transport
  processes: Scaling principles and asymptotic analysis\/}.
  \publ{Butterworth-Heinemann}.

\bibitem[Lemaire {\em et~al.\/}(2023)Lemaire, Blanc, Claudet, Gallier, Lobry \&
  Peters]{lemaire23}
{\sc \au{Lemaire, Elisabeth}, \au{Blanc, Fr{\'e}d{\'e}ric}, \au{Claudet,
  Cyrille}, \au{Gallier, Stany}, \au{Lobry, Laurent} \& \au{Peters, Fran{\c
  c}ois}} \yr{2023}  \at{Rheology of non-{Brownian} suspensions: a rough
  contact story}.  \jt{Rheologica Acta}  \bvol{62}~(5-6),  \pg{253--268}.

\bibitem[Miller {\em et~al.\/}(2009)Miller, Singh \& Morris]{miller09}
{\sc \au{Miller, Ryan~M.}, \au{Singh, John~P.} \& \au{Morris, Jeffrey~F.}}
  \yr{2009}  \at{Suspension flow modeling for general geometries}.
  \jt{Chemical Engineering Science}  \bvol{64}~(22),  \pg{4597--4610}.

\bibitem[Morris \& Brady(1998)]{morris98}
{\sc \au{Morris, J.F.} \& \au{Brady, J.F.}} \yr{1998}  \at{Pressure-driven flow
  of a suspension: {Buoyancy} effects}.  \jt{International Journal of
  Multiphase Flow}  \bvol{24}~(1),  \pg{105--130}.

\bibitem[Morris(2020)]{morris20a}
{\sc \au{Morris, Jeffrey~F.}} \yr{2020}  \at{Toward a fluid mechanics of
  suspensions}.  \jt{Physical Review Fluids}  \bvol{5}~(11).

\bibitem[Morris \& Boulay(1999)]{morris99}
{\sc \au{Morris, Jeffrey~F.} \& \au{Boulay, Fabienne}} \yr{1999}
  \at{Curvilinear flows of noncolloidal suspensions: {The} role of normal
  stresses}.  \jt{Journal of Rheology}  \bvol{43}~(5),  \pg{1213--1237}.

\bibitem[Morris \& Katyal(2002)]{morris02}
{\sc \au{Morris, Jeffrey~F.} \& \au{Katyal, Bhavana}} \yr{2002}
  \at{Microstructure from simulated {Brownian} suspension flows at large shear
  rate}.  \jt{Physics of Fluids}  \bvol{14}~(6),  \pg{1920--1937}.

\bibitem[Nadim \& Stone(1991)]{nadim91}
{\sc \au{Nadim, A.} \& \au{Stone, H.}} \yr{1991}  \at{The motion of small
  particles and droplets in quadratic flows}.  \jt{Studies in Applied
  Mathematics}  \bvol{85},  \pg{53--73}.

\bibitem[Noori {\em et~al.\/}(2024)Noori, Berry \& Harvie]{noori24a}
{\sc \au{Noori, Mohammad}, \au{Berry, Joseph~D.} \& \au{Harvie, Dalton~J.E.}}
  \yr{2024} Multifluid simulation of shear-induced migration in pressure-driven
  suspension flows. Submitted to arXiv 24/12/24.

\bibitem[Nott \& Brady(1994)]{nott94}
{\sc \au{Nott, P.} \& \au{Brady, J.}} \yr{1994}  \at{Pressure-driven flow of
  suspensions: simulation and theory}.  \jt{J. Fluid Mech.}  \bvol{275},
  \pg{157--199}.

\bibitem[Nott {\em et~al.\/}(2011)Nott, Guazzelli \& Pouliquen]{nott11}
{\sc \au{Nott, Prabhu~R.}, \au{Guazzelli, Elisabeth} \& \au{Pouliquen,
  Olivier}} \yr{2011}  \at{The suspension balance model revisited}.
  \jt{Physics of Fluids}  \bvol{23}~(4),  \pg{043304}.

\bibitem[Phan-Thien {\em et~al.\/}(1999)Phan-Thien, Fan \& Khoo]{phanthien99}
{\sc \au{Phan-Thien, N.}, \au{Fan, X.} \& \au{Khoo, B.}} \yr{1999}  \at{A new
  constitutive model for monodispersed suspensions of spheres at high
  concentrations}.  \jt{Rheol Act}  \bvol{38},  \pg{297--304}.

\bibitem[Phan‐Thien(1995)]{phanthien95a}
{\sc \au{Phan‐Thien, Nhan}} \yr{1995}  \at{Constitutive equation for
  concentrated suspensions in {Newtonian} liquids}.  \jt{Journal of Rheology}
  \bvol{39}~(4),  \pg{679--695}.

\bibitem[Phillips {\em et~al.\/}(1992)Phillips, Armstrong \& Brown]{phillips92}
{\sc \au{Phillips, R.}, \au{Armstrong, R.} \& \au{Brown, R.}} \yr{1992}  \at{A
  constitutive equation for concentrated suspensions that accounts for
  shear-induced particle migration}.  \jt{Phys. Fluids A}  \bvol{4}~(1),
  \pg{30--40}.

\bibitem[Rampall {\em et~al.\/}(1997)Rampall, Smart \& Leighton]{rampall97}
{\sc \au{Rampall, Indresh}, \au{Smart, Jeffrey~R.} \& \au{Leighton, David~T.}}
  \yr{1997}  \at{The influence of surface roughness on the particle-pair
  distribution function of dilute suspensions of non-colloidal spheres in
  simple shear flow}.  \jt{Journal of Fluid Mechanics}  \bvol{339},
  \pg{1--24}.

\bibitem[Wilson(2005)]{wilson05}
{\sc \au{Wilson, Helen~J.}} \yr{2005}  \at{An analytic form for the pair
  distribution function and rheology of a dilute suspension of rough spheres in
  plane strain flow}.  \jt{Journal of Fluid Mechanics}  \bvol{534},
  \pg{97--114}.

\bibitem[Wilson \& Davis(2000)]{wilson00}
{\sc \au{Wilson, Helen~J.} \& \au{Davis, Robert~H.}} \yr{2000}  \at{The
  viscosity of a dilute suspension of rough spheres}.  \jt{Journal of Fluid
  Mechanics}  \bvol{421},  \pg{339--367}.

\bibitem[Wilson \& Davis(2002)]{wilson02}
{\sc \au{Wilson, Helen~J.} \& \au{Davis, Robert~H.}} \yr{2002}  \at{Shear
  stress of a monolayer of rough spheres}.  \jt{Journal of Fluid Mechanics}
  \bvol{452},  \pg{425--441}.

\bibitem[Zhang \& Prosperetti(1997)]{zhang97}
{\sc \au{Zhang, D.~Z.} \& \au{Prosperetti, A.}} \yr{1997}  \at{Momentum and
  energy equations for disperse two-phase flows and their closure for dilute
  suspensions}.  \jt{Int. J. Multiphase Flow}  \bvol{23}~(3),  \pg{425--453}.

\bibitem[Zhang \& Acrivos(1994)]{zhang94}
{\sc \au{Zhang, K.} \& \au{Acrivos, A.}} \yr{1994}  \at{Viscous resuspension in
  fully developed laminar pipe flows}.  \jt{Int. J. Multiphase Flow}
  \bvol{20}~(3),  \pg{579--591}.

\bibitem[Zinchenko(1984)]{zinchenko84}
{\sc \au{Zinchenko, A.~Z.}} \yr{1984}  \at{Effect of hydrodynamic interactions
  between the particles on the rheological properties of dilute emulsions}.
  \jt{Journal of Applied Mathematics and Mechanics}  \bvol{48}~(2),
  \pg{198--206}.

\end{thebibliography}
\end{document}